\documentclass[]{interact}
\usepackage{geometry}
\usepackage[labelsep=period]{caption}

\usepackage{romannum}
\usepackage{multirow}

\usepackage{epstopdf}
\usepackage[caption=false]{subfig}

\usepackage[natbibapa,nodoi]{apacite}
\usepackage{hyperref}
\usepackage{lineno}
\usepackage{bbm}
\usepackage{url}
\usepackage{algorithm}
\usepackage{algorithmicx}
\usepackage{algpseudocode}
\usepackage{verbatim}

\usepackage{setspace}
\usepackage{breakurl}

\usepackage{hyperref}
\hypersetup{
  colorlinks=true,
  linkcolor=blue,
  citecolor=blue,
  urlcolor=blue}

\theoremstyle{definition}

\theoremstyle{remark}

\begin{document}


\title{Multi-objective probabilistic forecast combination for inventory demand}

\author{
\name{Shengjie Wang\textsuperscript{a}, Yanfei Kang\textsuperscript{a,$\ast$}, Evangelos Spiliotis\textsuperscript{b} and Fotios Petropoulos\textsuperscript{c}\thanks{Email: wsj19992017@buaa.edu.cn (Shengjie Wang), yanfeikang@buaa.edu.cn (Yanfei Kang, Corresponding author), spiliotis@fsu.gr (Evangelos Spiliotis), f.petropoulos@bath.ac.uk (Fotios Petropoulos)}}
\affil{\textsuperscript{a}School of Economics and Management, Beihang University, 37 Xueyuan Road, 100191, Beijing, China\\
\textsuperscript{b}Forecasting and Strategy Unit, School of Electrical and Computer Engineering, National Technical University of Athens, 9 Heroon Polytechniou Street, 15772 Zografos, Greece\\
\textsuperscript{c}School of Management, University of Bath, BA2 7AY, United Kingdom}
}

\maketitle

\pagenumbering{roman}

\begin{abstract}
Probabilistic forecasts are essential for inventory management, where decisions depend on the full distribution of future demand. While probabilistic forecast combination is widely used to improve statistical accuracy, most existing approaches optimize statistical loss alone and overlook operational objectives. However, in inventory settings, higher forecast accuracy does not necessarily translate into better decision performance, especially under nonlinear cost structures and multiple, potentially conflicting, decision targets. To address this gap, we propose a multi-objective probabilistic forecast combination framework that simultaneously considers forecast accuracy and inventory decision performance. The framework formulates forecast combination as a multi-objective optimization problem and derives a set of Pareto-optimal combinations, enabling explicit trade-offs between forecasting and operational goals. Empirical studies using Walmart retail data and Royal Air Force spare parts data demonstrate that the proposed approach achieves more balanced and robust performance than individual models, simple averaging, and single-objective optimization. Our results provide a practical and flexible framework for aligning probabilistic forecasting with inventory decision-making.

\vspace{10pt}

\noindent \textbf{PRACTITIONER SUMMARY}

\noindent Probabilistic forecast combination is an effective technique for improving forecasting performance. In the context of inventory management, this approach enables a more accurate description of future demand uncertainty. Recognizing that decision-makers are primarily concerned with how forecasts affect inventory performance, this study proposes a multi-objective probabilistic forecast combination framework that simultaneously optimizes both forecast accuracy and decision outcomes. Using real-world data from Walmart and Royal Air Force spare parts, we evaluate the proposed method from both forecasting and operational perspectives. The results show that our approach yields more accurate forecasts and more cost-effective decisions compared to individual models, simple averaging, and single-objective optimization. These findings suggest that integrating forecasting and operational objectives enhances the practical value of probabilistic forecast combination. For inventory managers, the proposed method offers a flexible framework for integrating demand forecast with inventory decision-making.

\noindent\textbf{Word count:} 7748
\end{abstract}


\clearpage   
\pagenumbering{arabic}

\section{Introduction}\label{intro}

Effective inventory management critically depends on how uncertainty in demand is quantified and translated into operational decisions. In many settings, replenishment policies—such as order-up-to levels or safety stock—are determined by specific quantiles of the demand distribution. This makes probabilistic forecasting, rather than point forecasting, central to modern inventory systems \citep{trapero2019quantile, kolassa2016evaluating}. Unlike point forecasts, which provide only the conditional expectation, probabilistic forecasts characterize the full predictive distribution, enabling decision-makers to account for the asymmetric trade-offs between holding costs and stockout penalties (e.g., in the newsvendor framework). Consequently, the economic value of a forecasting system is determined not only by its accuracy, but by how effectively its predictive distributions support operational decision rules under nonlinear cost structures \citep{THEODOROU2025414}.

Given the complexity of demand patterns, forecast combination has long been recognized as a robust strategy for improving predictive performance \citep{wang2023forecast}. By aggregating diverse component models, combinations mitigate the risks of model misspecification and overconfidence associated with individual forecasters \citep{hora2004probability}. Standard approaches to probabilistic forecast combination typically seek to minimize proper scoring rules—such as the Logarithmic Score (LogScore) or the Continuous Ranked Probability Score (CRPS)—thereby framing forecast combination as a statistical optimization problem \citep{hall2007combining, thorey2018ensemble, mcalinn2019dynamic}. While these methods yield calibrated predictive distributions, they implicitly assume that minimizing statistical loss aligns with minimizing inventory costs, which is not always the case \citep{SPILIOTIS2021108237}.

In inventory management, forecasting is not an end in itself. The ultimate goal is improved operational performance, typically measured by total inventory cost, service level, or stockout frequency \citep{goltsos2022inventory}. However, minimizing statistical loss does not necessarily translate into optimal operational outcomes, particularly under nonlinear decision rules such as the newsvendor model \citep{kourentzes2020optimising}. For instance, a forecast that minimizes the CRPS may fail to minimize total inventory costs or achieve desired service levels when loss functions are asymmetric or non-convex. This misalignment between forecast accuracy and decision performance has been increasingly recognized in research on contextual optimization and decision-focused learning \citep{sadana2025survey, elmachtoub2022smart}. Recent work seeks to bridge this gap by integrating prediction and decision-making within a unified framework \citep{qi2023practical, zhang2024optimal, olivares2024constructing}. More recently, decision-oriented objectives have also been incorporated into probabilistic forecast combination methods \citep{wang2024combining, stratigakos2025decision}.

Despite these advances, two important gaps remain from an operational perspective. First, most approaches optimize a single aggregate objective (e.g., total cost), overlooking the fact that inventory management involves multiple, often conflicting, performance criteria, such as cost, inventory level, and stockout frequency \citep{tsai2017simulation}. Second, existing methods frequently rely on rigid, composite loss functions or end-to-end architectures that obscure the trade-offs between statistical validity and operational performance. As a result, decision-makers are often constrained to implicit objective weightings, rather than being able to explicitly explore the trade-offs among competing goals. To the best of our knowledge, no existing approach jointly addresses probabilistic forecast combination and multi-objective inventory performance within a transparent Pareto-based framework.

Multi-objective optimization (MOO) provides a natural framework for addressing such competing criteria. Instead of aggregating multiple objectives into a single weighted metric, MOO characterizes the set of Pareto-efficient solutions, enabling explicit evaluation of trade-offs. In forecast combination, MOO has primarily been applied to point forecasts, particularly in energy forecasting applications \citep{wang2022novel, yang2024new, xing2024novel, chen2025novel}, where bias–variance trade-offs are emphasized. \citet{waychal2024adaptive} extend this approach to inventory-related objectives in point forecast combination. However, the application of MOO to probabilistic forecast combination—and, crucially, to the joint treatment of forecast accuracy and multiple inventory performance metrics—remains underdeveloped in the operations research literature.

To address this gap, we propose a multi-objective probabilistic forecast combination approach tailored for inventory management. We formulate the determination of combination weights as a MOO problem, in which forecast accuracy and multiple inventory performance metrics are treated as distinct, simultaneous objectives. Rather than collapsing these goals into a single scalar objective, the proposed approach constructs a Pareto frontier over combination weights. This enables retailers and supply chain planners to explicitly evaluate and select trade-offs that align with their operational strategies. The proposed framework is computationally tractable and can be implemented using standard multi-objective optimization techniques.

This study makes three key contributions. First, we propose a unified MOO framework that integrates probabilistic forecast combination with inventory decision-making, thereby explicitly addressing the misalignment between statistical scoring rules and operational performance. Second, we extend MOO to distributional forecast combination, allowing for the joint optimization of forecast accuracy and multiple inventory-related performance metrics. Third, we propose a flexible, model-agnostic procedure that can be applied with a wide class of probabilistic forecasting models, enhancing its applicability in real-world inventory settings. Empirical results demonstrate that the proposed approach improves inventory performance while revealing actionable trade-offs between cost and service level.

The rest of the paper is organized as follows. Section \ref{review} reviews the relevant literature on probabilistic forecast combination, contextual optimization methods in inventory management, and multi-objective forecast combination. Section \ref{proba-comb} presents the probabilistic forecasting methods and combination strategies. Section \ref{moo-comb} formulates the MOO problem and describes the solution methodology. Sections \ref{empirical} and \ref{case} present the empirical analysis and case studies, respectively. Finally, Section \ref{conclusion} concludes the paper.

\section{Related work}\label{review}

\subsection{Probabilistic forecast combination}

Research on probabilistic forecast combination remains active within the forecasting community \citep{wang2023forecast}. One of the most widely used approaches is linear pooling of individual forecasts, originally introduced by \citet{stone1961opinion}. A central challenge in linear pooling is the determination of combination weights. A straightforward solution is the simple average (arithmetic mean), as discussed in \citet{wallis2005combining} and \citet{o2006uncertain}. Due to its simplicity and empirical robustness, the simple average is often adopted as a benchmark for more sophisticated combination methods.

Motivated by developments in point forecast combination \citep{bates1969combination}, subsequent research has focused on deriving optimal weights through statistical optimization. \citet{hall2007combining} proposed minimizing the Kullback–Leibler information criterion (KLIC), which is equivalent to maximizing the logarithmic score. \citet{geweke2011optimal} established theoretical properties of logarithmic-score-based combinations. This line of work has been further extended to include efficient iterative optimization algorithms \citep{conflitti2015optimal}, time-varying weighting schemes \citep{kapetanios2015generalised,del2016dynamic}, and feature-based weights \citep{li2022bayesian}. 

Beyond the logarithmic score, the CRPS, a strictly proper scoring rule, has also been widely adopted as an optimization criterion \citep{raftery2005using, thorey2017online, thorey2018ensemble}. In certain applications, particularly in financial contexts, greater emphasis has been placed on tail performance, leading to scoring functions that prioritize extreme events \citep{opschoor2017combining, diebold2023aggregation}. Overall, this stream of research formulates probabilistic forecast combination as a statistical optimization problem aimed at improving forecast accuracy under proper scoring rules, with limited explicit consideration of downstream decision performance.

In addition to linear pooling, several alternative combination frameworks have been proposed. Nonlinear pooling methods \citep{gneiting2013combining, ranjan2010combining, van2024crps} aim to recalibrate combined distributions to address potential deficiencies of linear pooling, although their empirical gains are often modest \citep{baran2018combining}. Bayesian Model Averaging (BMA) provides a formal treatment of model uncertainty \citep{garratt2003forecast}, but relies on the assumption that the true data-generating process lies within the candidate model set \citep{wright2008bayesian}. Bayesian Predictive Synthesis \citep[BPS;][]{mcalinn2019dynamic}, together with its multivariate \citep{mcalinn2020multivariate} and decision-oriented extensions \citep{tallman2024bayesian}, accounts for dependencies among forecast distributions. Quantile-based combinations \citep{lichtendahl2013better, trapero2019quantile, li2022bayesian} provide another flexible alternative, while angular combination \citep{taylor2026angular} bridges probability and quantile pooling. Comprehensive reviews of the field can be found in \citet{wang2023forecast}.

Most of the literature described above focuses on statistical properties and forecasting performance. More recently, decision-oriented perspectives have begun to emerge. \citet{tallman2024bayesian} incorporate decision outcomes within a Bayesian predictive synthesis framework. \citet{wang2024combining} propose optimizing combination weights with respect to inventory cost functions. \citet{stratigakos2025decision} introduce decision-focused linear pooling to explicitly align probabilistic forecast combination with operational objectives. These developments highlight the growing recognition that probabilistic forecast combination should be evaluated not only in terms of statistical accuracy but also in terms of economic value. However, existing approaches predominantly rely on single-objective formulations and do not explicitly capture trade-offs across multiple operational criteria, thereby motivating further research at the interface of forecasting and operations.

\subsection{Contextual optimization methods in inventory management}

Research on integrating prediction with decision optimization has expanded rapidly in recent years. \citet{sadana2025survey} provide a comprehensive review under the framework of contextual optimization, which studies how predictive information can be incorporated directly into decision-making problems. They categorize the literature into three main approaches: decision rule optimization, sequential learning and optimization (SLO), and integrated learning and optimization (ILO). In this subsection, we focus on developments within inventory management. 

Decision rule optimization constructs a direct mapping from observed data to decision policies, bypassing an explicit forecasting stage. For example, \citet{huber2019data} propose a machine-learning-based solution to the newsvendor problem using quantile regression. \citet{oroojlooyjadid2020applying} employ deep neural networks to determine order quantities without specifying a parametric demand distribution. \citet{bertsimas2023dynamic} combine machine learning with robust optimization in multi-period decision problems. Similarly, \citet{zhang2024optimal} derive feature-based ordering policies within a distributionally robust optimization framework. These approaches directly optimize operational objectives but typically do not produce explicit probabilistic forecasts, thereby limiting their applicability in settings where full predictive distributions are required for decision-making.

SLO decouples prediction and decision stages. A forecasting model is first trained to estimate demand or contextual distributions, which are then used as inputs to an optimization model. For instance, \citet{ferreira2016analytics} forecast daily sales using tree-based models and subsequently optimize pricing decisions. \citet{lin2022data} estimate similarity between products and incorporate it into a correlation-weighted inventory objective. SLO preserves modularity between forecasting and optimization but does not guarantee alignment between forecast loss and decision performance, particularly when evaluation metrics differ across stages, which may lead to suboptimal operational decisions.

ILO combines prediction and decision tasks into a single optimization problem, directly targeting decision performance. This formulation often requires differentiable or surrogate objectives to enable gradient-based optimization \citep{elmachtoub2022smart}. Representative examples include the end-to-end frameworks of \citet{NIPS2017_3fc2c60b} and \citet{qi2023practical}. To alleviate computational complexity, \citet{olivares2024constructing} approximate the integrated objective using linearization techniques. While ILO explicitly aligns learning with operational objectives, it typically results in more complex optimization problems and may require customized model structures, potentially limiting scalability and ease of implementation in real-world inventory systems.

Among these approaches, SLO and ILO are most closely related to our study because they incorporate both forecasting and decision components. However, SLO may suffer from forecast–decision misalignment, as predictive models are trained independently of operational objectives. In contrast, ILO directly optimizes decision performance but at the cost of increased computational and modeling complexity. Both paradigms generally treat decision performance as the ultimate objective and often focus on a single operational criterion. This motivates the need for a flexible framework that can accommodate probabilistic forecasts while simultaneously handling multiple decision objectives within an optimization-based combination structure, while preserving transparency and interpretability of trade-offs for decision-makers.

\subsection{Forecast combination based on multi-objective optimization}

MOO addresses problems involving multiple, potentially conflicting objectives. Two main approaches are commonly adopted. The first seeks to identify the set of Pareto-optimal solutions, thereby characterizing the trade-off frontier among objectives. The second aggregates multiple objectives into a single composite function, typically through weighted summation, and solves a scalarized optimization problem. Both approaches have been explored in the context of forecast combination.

The Pareto-based approach generally relies on heuristic or evolutionary algorithms to approximate the non-dominated solution set. It has been widely applied in energy forecasting. For example, \citet{wang2022novel} employed a gray wolf algorithm to balance MAPE and the standard error of forecast errors. \citet{yang2024new} combined interval forecasts by minimizing both the mean and variability of interval widths using the Archimedes optimization algorithm. Similarly, \citet{chen2025novel} forecasted carbon prices using a football team training algorithm to jointly minimize the mean and standard deviation of forecast errors. These studies mainly focus on statistical trade-offs—particularly bias–variance or accuracy–stability trade-offs—while decision-oriented objectives receive limited attention. Moreover, most applications concentrate on point or interval forecasts, and the extension of Pareto-based MOO to probabilistic forecast combination remains relatively underexplored, especially in settings where operational decisions depend on full predictive distributions, such as inventory control problems.

The scalarization approach transforms a multi-objective problem into a single-objective one by assigning weights to different criteria. \citet{waychal2024adaptive} incorporate both forecasting accuracy and inventory-related objectives through weighted aggregation based on decision-maker preferences. Similarly, the regret objective with CRPS regularization proposed by \citet{stratigakos2025decision} can be interpreted as a weighted composite objective. While scalarization simplifies computation, it requires specifying relative importance across objectives in advance, introducing subjectivity into the optimization process and potentially limiting the exploration of the full trade-off structure, particularly when preferences are uncertain or context-dependent, or when decision-makers seek to explicitly evaluate alternative trade-offs.

Overall, existing multi-objective forecast combination methods either emphasize statistical trade-offs or rely on subjective aggregation of objectives. The joint consideration of probabilistic forecast quality and multiple operational decision metrics within a unified Pareto-based framework remains insufficiently developed, particularly in inventory management settings, thereby motivating the approach proposed in this study, which explicitly characterizes and exploits these trade-offs within a probabilistic forecast combination framework.

\section{Probabilistic forecast combination}\label{proba-comb}

\subsection{Combination formulation}

We consider $K$ individual probabilistic forecasts $F_1^{(t)}(y),\cdots, F_K^{(t)}(y)$, where $F_i^{(t)}(y)$ represents the $i$-th cumulative distribution function (CDF) forecast for time series ${y}$ at period $t$. To aggregate these individual predictive distributions, we employ the linear pool method (also known as the linear opinion pool), a standard approach in probabilistic forecasting~\citep{hall2007combining}. The combined forecast is defined as
\begin{equation}
    F_{\text{comb}}^{(t)}(y)=\sum_{i=1}^{K} w_iF_i^{(t)}(y),
    \label{eq:comb}
\end{equation}
where $\mathbf{w} = (w_1, w_2, \cdots, w_K)$  is the vector of combination weights.
Equation~\eqref{eq:comb} defines a finite linear mixture of predictive distributions. 
To ensure that $F_{\text{comb}}^{(t)}(y)$ is a valid CDF, the weights $w_i$ are restricted to be non-negative and sum to one. For tractability, we assume that the weights are time-invariant over the forecasting horizon, although the proposed framework can be extended to allow for time-varying weights.

For the linear pooling method, predictive performance depends critically on the specification of the weights. A simple and widely used approach is to assign equal weights, i.e., the arithmetic mean (simple average), which serves as a standard benchmark in forecast combination studies. Alternatively, the weights can be obtained by solving optimization problems under different objective functions, depending on the desired trade-offs and objectives. These optimization-based weighting schemes will be discussed in Section~\ref{moo-comb}.

\subsection{Individual forecasts}

Given the prevalence of count data in inventory management, particularly in the case of intermittent demand, this study explores four methods/models that generate discrete distributions. The first two methods are bootstrap-based, while the remaining two are distribution-based models. 

\subsubsection*{Bootstrap methods}

Bootstrap methods generate forecasts by resampling historical data. For count data in inventory management, we consider two bootstrap methods from \citet{willemain2004new} and \citet{zhou2011comparison}, referred to as WSS and ZV in this paper. The WSS method estimates the probability and magnitude of positive demand, and the forecast distribution is constructed by resampling both the occurrence of demand and the size of positive demand based on these estimates. In contrast, the ZV method resamples the distribution of the time intervals between two positive demands and the size of positive demand to derive the forecast distribution. For forecasting, we resample historical data 1,000 times to obtain a Monte Carlo approximation of the predictive distribution.

\subsubsection*{Distribution-based models}

Distribution-based models forecast discrete demand distributions based on specific statistical models, such as the Poisson and negative binomial distributions. In the presence of time-varying demand, we adopt the damped dynamic model from \citet{snyder2012forecasting}, which models the mean of the distribution as:
\begin{equation}
    \mu_t=(1-\phi-\alpha)\mu+\phi\mu_{t-1}+\alpha y_{t-1},
    \label{eq:distribution model}
\end{equation}
where the lagged mean coefficient $\phi$, the lagged observation coefficient $\alpha$, and the long-run mean $\mu$ are positive, and satisfy $\phi+\alpha<1$. $y_t$ is the demand at period $t$. As the means vary over time, the distributional parameters evolve accordingly; further model details can be found in \citet{snyder2012forecasting}. Maximum likelihood estimation (MLE) can be applied to estimate the model parameters, followed by the generation of predictive distributions.

\section{Multi-objective-optimization-based combination}\label{moo-comb}

In Section~\ref{proba-comb}, we discussed that the weights in forecast combination can be determined through optimization. A central question, therefore, concerns the choice of optimization objectives. In the probabilistic forecasting literature, combination weights are typically selected to improve predictive accuracy under proper scoring rules, such as the logarithmic score \citep{conflitti2015optimal} and the CRPS \citep{thorey2018ensemble}. 

However, statistical optimality does not necessarily translate into decision optimality. In inventory management, minimizing forecast error alone does not necessarily lead to lower inventory costs \citep{kourentzes2020optimising}. From an operational perspective, decision metrics—such as total cost, inventory holding levels, and stockout quantity—are often more relevant than abstract statistical scores. Furthermore, operational decision-making is rarely driven by a single goal; it typically involves balancing conflicting objectives, such as the trade-off between holding costs and service levels. Consequently, an effective optimization framework must explicitly accommodate these competing criteria. This naturally motivates the formulation of forecast combination as a MOO problem. In this section, we define the relevant objective functions and formulate the probabilistic forecast combination as a MOO problem.

\subsection{Optimization objectives}

We consider four distinct objective functions categorized into two groups: a statistical forecast metric and three inventory decision metrics.

\subsubsection*{Forecast evaluation metric}

To evaluate statistical accuracy, we employ the Distributional Ranked Probability Score (DRPS), which measures the divergence between the predictive cumulative distribution function (CDF) and the observed value. For a count data setting, the DRPS is defined as:

\begin{equation}
    \text{DRPS}(F_{\text{comb}}^{(t)}(y),y_t)=\sum_{n=0}^{+\infty} (F_{\text{comb}}^{(t)}(n)-\boldsymbol{1}[y_{t} \leq n])^2,
    \label{eq:DRPS}
\end{equation}

where $F_{\text{comb}}^{(t)}(n)$ denotes the CDF of the combined forecast evaluated at value $n$, and $\boldsymbol{1}[\cdot]$ represents the indicator function. The DRPS is a strictly proper scoring rule widely used for evaluating probabilistic forecasts of count data~\citep{snyder2012forecasting, wang2024combining}, ensuring that the true data-generating distribution is optimal in expectation.

\subsubsection*{Decision evaluation metrics}

We consider a single-period inventory setting where the decision maker observes sales in period $t$ and generates a probabilistic forecast for period $t+1$. Replenishment decisions are finalized prior to the start of period $t+1$. The decision maker specifies a target service level $\tau \in (0,1)$ and sets the order-up-to level as the $\tau$-quantile of the forecast distribution. Inventory performance is evaluated using three metrics: total cost, average holding stock, and stockout quantity. These are defined over an evaluation window of length $h$ as follows:

\begin{equation}
\begin{aligned}
    \text{Cost}
    &= c_1 \cdot \text{Holding}
       + c_2 \cdot \text{Stockout}, \\
    \text{Holding}
    &= \frac{1}{h} \sum_{t=T-h+1}^{T}
       \big[ q_t(\tau) - y_t \big]_+, \\
    \text{Stockout}
    &= \frac{1}{h} \sum_{t=T-h+1}^{T}
       \big[ y_t - q_t(\tau) \big]_+,
\end{aligned}
\label{eq:inventory-metrics}
\end{equation}

where $c_1$ and $c_2$ denote the unit holding cost and unit stockout cost, respectively. 
Consistent with the classical newsvendor model, the optimal target service level corresponds to the critical ratio $\tau = \frac{c_2}{c_1 + c_2}$~\citep{huber2019data}. $q_t(\tau)$ is the order-up-to level, defined as the $\tau$-quantile of the forecast distribution (either combined or individual), linking probabilistic forecasts directly to operational decisions.

\subsubsection*{Single-objective optimization}

To provide a baseline for comparison, we first consider two single-objective formulations: maximizing statistical accuracy (DRPS-opt) and minimizing operational cost (Cost-opt). The DRPS-opt problem minimizes the average statistical score:

\begin{equation}
\begin{aligned}
\min_{\mathbf{w}} \quad  &\frac{1}{h}\sum_{t=T-h+1}^{T}\text{DRPS}(F_{\text{comb}}^{(t)}(y), y_t), \\
\text{s.t.} \quad &  \mathbf{w}^\top \mathbf{1} = 1,\\
\quad & \mathbf{w} \geq 0.&
\end{aligned}
\label{eq:drps-opt}
\end{equation}
As the DRPS is quadratic in the weight vector $\mathbf{w}$, Problem~(\ref{eq:drps-opt}) can be solved efficiently as a quadratic programming problem with linear constraints.
The Cost-opt problem minimizes the total inventory cost:
\begin{equation}
\begin{aligned}
\min_{\mathbf{w}} \quad  &\text{Cost}, \\
\text{s.t.} \quad  & \mathbf{w}^\top \mathbf{1} = 1, \\
\quad &\mathbf{w} \geq 0.
\end{aligned}
\label{eq:cost-opt}
\end{equation}

Solving Problem~(\ref{eq:cost-opt}) presents computational challenges. Because the forecast distributions are discrete (as described in Section~\ref{proba-comb}), the mapping from weights to quantiles—and subsequently to costs—introduces step functions, rendering the objective function non-smooth and non-convex. Consequently, gradient-based methods are not applicable. Following \citet{wang2024combining}, we employ Particle Swarm Optimization~\citep[PSO; ][]{kennedy1995particle} to solve this problem, as it is well-suited for non-convex and non-differentiable optimization landscapes.

\subsection{Multi-objective optimization}\label{moo}

As noted by \citet{wang2024combining}, statistical and operational objectives often conflict; minimizing statistical error does not imply minimal cost. To explicitly address these trade-offs, we formulate forecast combination as a MOO problem:
\begin{equation}
\begin{aligned}
\min_{\mathbf{w}} \quad &
\mathbf{f}(\mathbf{w}) 
= \big( e_1(\mathbf{w}, \mathcal{F}, \mathbf{y}), \ldots, e_m(\mathbf{w}, \mathcal{F}, \mathbf{y}); 
u_1(\mathbf{w}, \mathcal{F}, \mathbf{y}), \ldots, u_l(\mathbf{w}, \mathcal{F}, \mathbf{y}) \big), \\
\text{s.t.} \quad &
 g(\mathbf{w}) \le 0, \\&
 h(\mathbf{w}) = 0.
\end{aligned}
\label{eq:moo-comb-problem}
\end{equation}
where $\mathbf{w} = (w_1, \ldots, w_K)$ denotes the weight vector, 
$\mathbf{y} = (y_{T-h+1}, \ldots, y_T)$ represents the realized demand over the validation horizon of length $h$, and 
$\mathcal{F} = \big( F_1^{(t)}(y), \ldots, F_K^{(t)}(y) \big)_{t = T-h+1}^{T}$ denotes the set of component probabilistic forecasts over this horizon. 
The objective vector consists of $m$ statistical error metrics (denoted by $e_j$) and $l$ decision utility metrics (denoted by $u_k$), thereby jointly capturing predictive accuracy and operational performance.

In this study, we implement this framework using two specific configurations. The first configuration considers a bi-objective problem balancing forecast accuracy against total cost, as formulated in Equation~(\ref{eq:2-obj}):
\begin{equation}
\min_{\mathbf{w}}\quad  \left[
e_1 = \frac{1}{h}\sum_{t=T-h+1}^{T}\text{DRPS}(F_{\text{comb}}^{(t)}(y),y_t), \quad
u_1 = \text{Cost}
\right].
\label{eq:2-obj}
\end{equation}
The second configuration decomposes the cost metric to explicitly manage the trade-off between overstocking and understocking, resulting in the tri-objective formulation in Equation~(\ref{eq:3-obj}):
\begin{equation}
\min_{\mathbf{w}} \quad \left[
e_1 = \frac{1}{h}\sum_{t=T-h+1}^{T}\text{DRPS}(F_{\text{comb}}^{(t)}(y),y_t), \quad
u_1 = \text{Holding}, \quad
u_2 = \text{Stockout}
\right].
\label{eq:3-obj}
\end{equation}

The constraints $g(\mathbf{w}) \leq 0$ and  $h(\mathbf{w}) = 0$ ensure the validity of the linear pool. We enforce the standard constraints $w_i \geq 0, \quad i=1,\cdots,K$ and $\sum_{i=1}^Kw_i=1$ in our setting. This formulation transforms the selection of combination weights into a MOO problem, allowing for the generation of a Pareto-optimal set of solutions that reflect varying trade-offs between forecast accuracy and inventory performance, rather than imposing a single pre-specified preference structure.

\subsubsection*{Pareto optimality and solution approaches}

Unlike single-objective optimization, MOO problems rarely admit a single global solution that minimizes all objectives simultaneously. Instead, the goal is to identify a set of solutions representing the optimal trade-offs between conflicting objectives. The primary criterion for comparing solutions in this context is Pareto dominance.
Consider a decision space $\Omega$ and a vector of $m$ objective functions $\mathbf{F}(x) = (f_1(x), \dots, f_m(x))$. For two solutions $x_A, x_B \in \Omega$, we state that $x_A$ Pareto dominates $x_B$ (denoted as $x_A \succ x_B$) if and only if:
\begin{equation}
    \left(\forall i \in \{1,\dots,m\}: f_i(x_A) \leq f_i(x_B)\right) \wedge \left(\exists j \in \{1,\dots,m\}: f_j(x_A) < f_j(x_B)\right).
\end{equation}
In other words, $x_A$ is strictly superior to $x_B$ in at least one objective without being inferior in any other. A solution $x^*$ is termed Pareto-optimal (or non-dominated) if there exists no $x \in \Omega$ such that $x \succ x^*$. The set of all such solutions in the decision space is the Pareto set ($\mathcal{P}_S$), and the corresponding image in the objective space is the Pareto frontier ($\mathcal{P}_F$).

Broadly, there are two strategies to solve MOO problems like Equation~\eqref{eq:moo-comb-problem}. The first aggregates multiple objectives into a single composite metric (typically a weighted sum), thereby converting the multi-objective problem into a single-objective one \citep[e.g.,][]{waychal2024adaptive}. The second strategy aims to approximate the entire Pareto set directly \citep[e.g.,][]{yang2024new}.

In this study, we adopt the second approach for two reasons. First, the aggregation approach requires the a priori assignment of weights to each objective, which introduces subjectivity before the actual trade-offs are known. In contrast, finding the Pareto set decouples the optimization process from the decision-making process; the decision maker selects the most appropriate solution only after observing the available trade-offs. Second, if the Pareto front is non-convex, weighted-sum methods are theoretically incapable of identifying solutions in the non-convex regions. Direct MOO can identify these solutions, offering a more comprehensive view of the performance landscape, which is particularly important in inventory settings with nonlinear cost structures.

\subsubsection*{Optimization algorithm}

Finding the Pareto set for a MOO problem presents a significant challenge. Typically, heuristic algorithms are employed for this purpose. In this paper, we utilize the Non-dominated Sorting Genetic Algorithm \Romannum{3} \citep[NSGA-\Romannum{3};][]{deb2013evolutionary, jain2013evolutionary}, a well-established algorithm designed to handle multiple objectives efficiently. NSGA-\Romannum{3} operates by evolving a population of candidate solutions over successive generations. It balances two competing goals: minimizing distance to the Pareto front (convergence) and maximizing the spread of solutions across the front (diversity). To achieve this, the algorithm employs non-dominated sorting to stratify the population into hierarchical dominance layers, ensuring that superior solutions survive. In addition, NSGA-\Romannum{3} maintains population diversity through reference-point approaches, which guides the selection process to ensure a uniform distribution of solutions along the trade-off surface. This makes it particularly robust for problems with three or more objectives compared to its predecessor, NSGA-\Romannum{2}.

We implement the optimization procedure using the \texttt{pymoo} framework in Python \citep{pymoo}. While the development of novel MOO algorithms remains an active and rapidly evolving field \citep[e.g.,][]{hao2024constrained}, the primary contribution of this work lies in the methodological framework for probabilistic forecast combination rather than algorithmic development. Therefore, we select NSGA-\Romannum{3} as a representative and reliable solver to demonstrate the efficacy of our approach, while maintaining computational tractability. Future research may explore the comparative performance of alternative optimization heuristics within this specific inventory management context.

\subsubsection*{Selection from the Pareto set}

Selecting a single actionable policy from the Pareto set involves trading off competing objectives. To minimize subjectivity and ensure scalability, we implement two automated selection criteria: the ideal-point method and the performance-index method.

\paragraph*{Ideal-point method}
Adapted from \citet{ganjehkaviri2017genetic}, this geometric approach seeks the solution closest to a theoretical ``utopia'' point where all objectives are simultaneously minimized. We employ a three-step procedure: (1) we apply min-max normalization to standardize the objective scales, with a modification to handle zero values common in inventory metrics (e.g., zero stock-outs); (2) we calculate the Euclidean distance of each Pareto-optimal solution from the origin (representing the ideal point) in the normalized space; and (3) we select the solution that minimizes this distance, thereby identifying a balanced compromise across all objectives.

\paragraph*{Performance-index method}
Inspired by the robust optimization framework of \citet{weber2025relatively}, this method prioritizes solutions that perform well relative to their worst-case potential. For a minimization problem, we define the performance index $\rho(x)$ as the minimum ratio of the best possible value to the actual value across all objectives:
\begin{equation}
    \rho(x) = \min_{i \in \Omega} \frac{f_i^{\min}}{f_i(x)},
\end{equation}
where $\Omega$ is the set of objective indices and $f_i^{\min} = \min\limits_{y \in \mathcal{P}} f_i(y)$ is the minimum value for the $i$-th objective within the Pareto set. This index $\rho(x) \in [0,1]$ identifies the ``weakest link'' among objectives; we seek the solution that maximizes this robustness metric, thereby favoring solutions with well-balanced performance across all criteria.

The target set for this method is the intersection of the maximum performance set and the Pareto set, denoted $\mathcal{R} = \arg\max\limits_x \rho(x) \cap \mathcal{P}$. While \citet{weber2025relatively} proves $\mathcal{R}$ is non-empty for continuous functions, the property also holds for our discrete domain. The logic is straightforward: if a solution $x'$ maximizes $\rho(x)$ but is not Pareto optimal ($x' \notin \mathcal{P}$), there must exist a dominating solution $x'' \in \mathcal{P}$ such that $f_i(x'') \leq f_i(x')$ for all $i$. This implies $\rho(x'') \geq \rho(x')$. Since $x'$ is maximal, $x''$ must also be maximal and lies within $\mathcal{P}$. Thus, a valid, robust solution is guaranteed to exist within the set identified by NSGA-\Romannum{3}, ensuring consistency between robustness and Pareto efficiency.

\section{Empirical evaluation: Walmart data}\label{empirical}

\subsection{Data description and forecasting process}

We apply the MOO-based forecast combination framework to the M5 dataset \citep{makridakis2022m5}, which contains actual sales data from Walmart stores in the United States. The dataset comprises 30,490 daily time series representing 3,049 products across 10 stores over 1,941 days. This dataset exemplifies intermittent demand patterns, with approximately 60.1\% zero observations, making it particularly challenging for probabilistic forecasting.

\begin{figure}
    \centering
    \includegraphics[width=\linewidth]{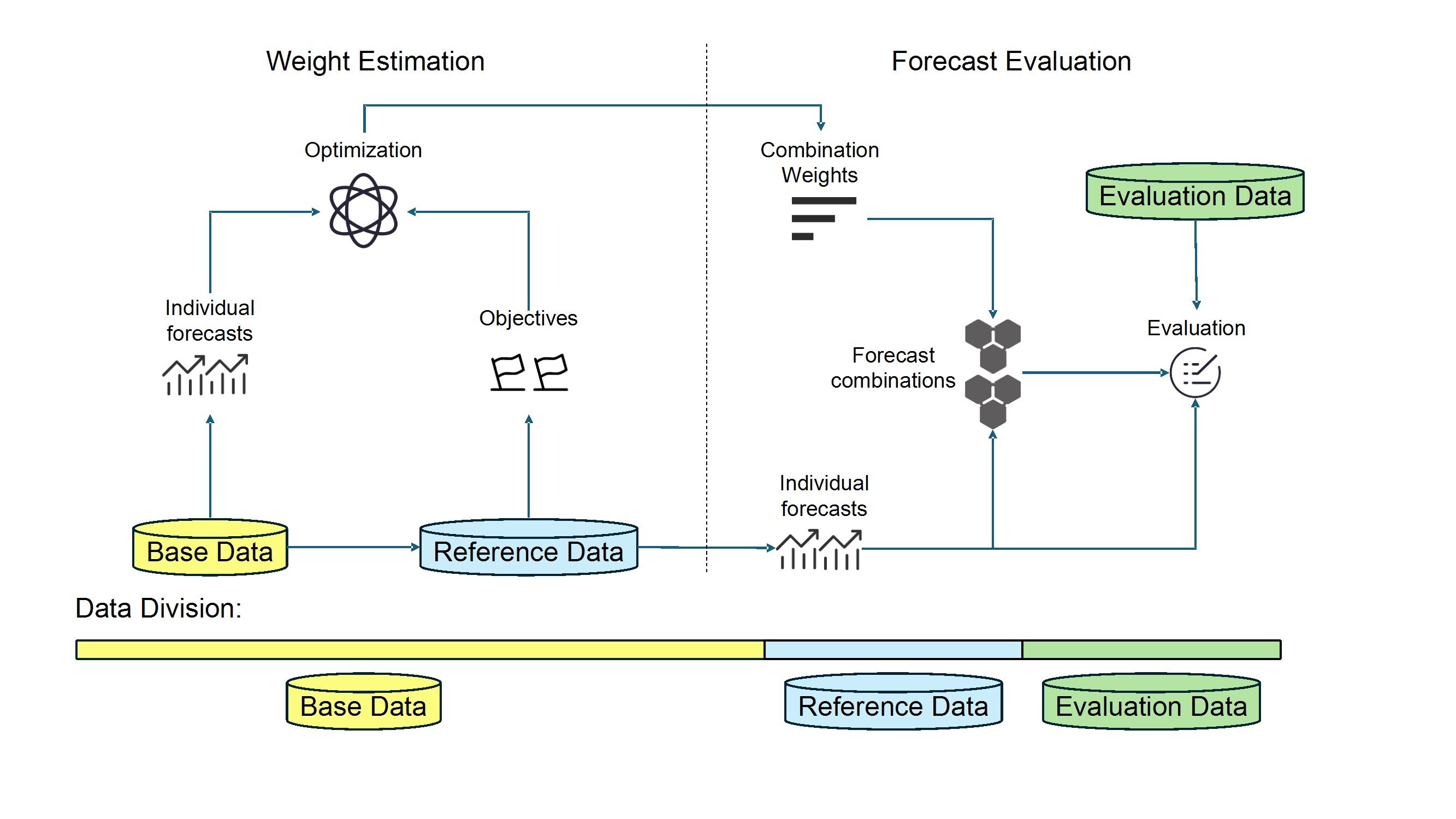}
    \caption{The process of multi-objective probabilistic forecast combination, including weight estimation and forecast evaluation.}
    \label{fig:process}
\end{figure}

\paragraph*{Experimental setup}
Figure~\ref{fig:process} illustrates our forecasting procedure. Each time series is partitioned into three sequential segments. The base data is used to generate individual forecasts from competing methods. The reference data serves to estimate optimal combination weights and produce combined forecasts for the evaluation data, where forecast accuracy and decision quality are assessed. Notably, individual forecasts for the evaluation period are produced using both base and reference data concatenated, ensuring fair comparison with the combined forecasts, while avoiding information leakage in weight estimation. Following the M5 competition protocol, we forecast a 28-day horizon. The data partition is structured as follows: base data spans from each series' first positive demand observation to day 1857 (57 days before the end); reference data covers days 1858--1913 (the penultimate 56 days); and evaluation data comprises the final 28 days (days 1914--1941). This rolling-window structure reflects realistic forecast production environments where recent data informs combination weights before out-of-sample evaluation, mimicking an operational rolling forecast setting.

\paragraph*{Cost configurations and optimization settings}
We evaluate three inventory cost scenarios corresponding to different target service levels. The unit holding cost is fixed at $c_1 = 1$, while the unit stockout cost $c_2$ takes values 4, 9, and 19, yielding target service levels $\tau$ of 80\%, 90\%, and 95\%, respectively. These settings span a realistic range of inventory policies, from lower-margin products (80\% service level) to critical high-service items (95\% service level).
For the optimization problems defined in Equations~\eqref{eq:drps-opt}, \eqref{eq:cost-opt}, \eqref{eq:2-obj}, and \eqref{eq:3-obj}, we apply cost pairs $(c_1, c_2) \in \{(1,4), (1,9), (1,19)\}$ and service levels $\tau \in \{0.8, 0.9, 0.95\}$ as appropriate. Evaluation metrics align with optimization objectives to ensure consistency between training and testing. We denote combinations based on Problems~\eqref{eq:drps-opt}, \eqref{eq:cost-opt}, \eqref{eq:2-obj}, and \eqref{eq:3-obj} as DRPS-opt, Cost-opt, NSGA-\Romannum{3}-c, and NSGA-\Romannum{3}-hs, respectively. Additionally, we include the simple average (SA) as a baseline combination method.

\paragraph*{Data preprocessing}
A total of 1,587 series exhibit exclusively zero demand during the reference period while recording positive demand in the evaluation period. Including these series would bias weight estimation toward methods producing zero forecasts, yielding artificially inflated errors. Moreover, such patterns likely represent product reinstatements or promotional events that cannot be reliably forecast from historical sales alone. Evaluating forecast performance on these series would be inappropriate, as the information required for accurate prediction is absent from the available data. Consequently, we exclude these 1,587 series from analysis, retaining 28,903 series for evaluation. This filtering ensures that our results reflect forecast performance under realistic conditions where historical patterns contain predictive information, rather than structural breaks or unobserved external interventions.

\subsection{Main performance results}\label{m5-result}

This subsection evaluates the performance of our proposed methods using the ideal-point selection criterion. As demonstrated in the appendix, this criterion offers greater stability compared to the performance-index method; therefore, it is adopted for the remainder of this analysis. Tables \ref{tab:single-80} through \ref{tab:single-95} present the forecast and decision metrics, while Table \ref{tab:single-rank} summarizes the average rankings of the four metrics across three scenarios. In all tables, superior performance is highlighted in \textbf{bold}, and the second best results are indicated in \textit{italics}, with rankings computed separately for each cost configuration to ensure comparability across service levels.

\begin{table}[!ht]
\centering
\caption{Forecast and decision metrics for M5 data with cost parameters $(c_1,c_2)=(1,4)$. DRPS assesses forecast accuracy, and the other three evaluate decision quality. \textbf{Bold} numbers highlight the superior performance, while \textit{italics} ones indicate the second best results.}
\label{tab:single-80}
{\begin{tabular}{llccccc}
\toprule
Type & Method & DRPS & Cost & Holding & Stockout \\
\midrule 
\multirow{4}*{Individuals} & WSS & 1.3363 & 73.5293 & 34.6778& 9.7129 \\
~ & ZV  & 1.1760 & 70.2831 & 39.3137& \textit{7.7423} \\
~ & POIS  & \textbf{1.1290} & \textbf{64.5532} & \textbf{30.4065}&  8.5367\\
~ & NB  &1.2350 & 67.1243 & 32.8663& 8.5645 \\
\hline
\multirow{5}*{Combinations} & SA & 1.1860& 65.6164 & 32.7365 & 8.0368\\
~ & NSGA-\Romannum{3}-c  & \textit{1.1419} & 65.3958 & 33.5079&  7.9720\\
~ & NSGA-\Romannum{3}-hs  & 1.1427 & 65.7930 & 33.5842& 8.0522\\
~ &  Cost-opt & 1.1953 & \textit{65.1533} & \textit{32.1518}& 8.2504\\
~ & DRPS-opt & 1.1605 & 65.4852 & 35.2844& \textbf{7.5502} \\
\bottomrule 
\end{tabular}}
\end{table}

\begin{table}[!ht]
\centering
\caption{Forecast and decision metrics for M5 data with cost parameters $(c_1,c_2)=(1,9)$. \textbf{Bold} highlights the best performance, while \textit{italics} indicates the second best results.}
\label{tab:single-90}
{\begin{tabular}{llccccc}
\toprule
Type & Method & DRPS & Cost & Holding & Stockout \\
\midrule 
\multirow{4}*{Individuals} & WSS & 1.3363 & 104.7048& 69.8816& 3.8693 \\
~ & ZV  & 1.1760 & 100.4582 & 66.9763& \textbf{3.7202} \\
~ & POIS  & \textit{1.1290} & 93.8980 & \textbf{46.1036}&  5.3105\\
~ & NB  &1.2350 & 97.5008 & 63.9150& 3.7318\\
\hline
\multirow{5}*{Combinations} & SA & 1.1860& 93.9408 & 60.4431 & 8.0368\\
~ & NSGA-\Romannum{3}-c  & 1.1337 & \textit{92.3667} & 55.6951&  4.0746\\
~ & NSGA-\Romannum{3}-hs  & \textbf{1.1263} & 92.4628 & 54.9600& 4.1670\\
~ &  Cost-opt & 1.6677 & \textbf{92.2898} & \textit{54.8307}& 4.1621\\
~ & DRPS-opt & 1.1605 & 93.8811 & 57.7879& \textit{4.0103} \\
\bottomrule 
\end{tabular}}
\end{table}

\begin{table}[!ht]
\centering
\caption{Forecast and decision metrics for M5 data with cost parameters $(c_1,c_2)=(1,19)$. \textbf{Bold} highlights the best performance, while \textit{italics} indicates the second best results.}
\label{tab:single-95}
{\begin{tabular}{llccccc}
\toprule
Type & Method & DRPS & Cost & Holding & Stockout \\
\midrule 
\multirow{4}*{Individuals} & WSS & 1.3363 &139.6664 & 110.3595& \textbf{1.5425} \\
~ & ZV  & 1.1760 & 132.5307 & 96.9735& 1.8714 \\
~ & POIS  & \textit{1.1290} &129.134 & \textbf{61.1750}&  3.5768\\
~ & NB  &1.2350 & 132.3501 & 99.0543& 1.7524 \\
\hline
\multirow{5}*{Combinations} & SA & 1.1860& 124.5485& 91.3683 & \textit{1.7463}\\
~ & NSGA-\Romannum{3}-c  & 1.1351 & 122.2545 & 79.3229&  2.2596\\
~ & NSGA-\Romannum{3}-hs  & \textbf{1.1230} & \textit{121.8918} & \textit{76.0481}& 2.4128\\
~ &  Cost-opt & 1.7506 & \textbf{121.6063} & 77.6393& 2.3141\\
~ & DRPS-opt & 1.1605 & 124.8574& 82.3413& 2.2377 \\
\bottomrule 
\end{tabular}}
\end{table}

\begin{table}[!ht]
\centering
\caption{Average ranks of forecast and decision performance in three cost scenarios for M5 data.}
\label{tab:single-rank}
{\begin{tabular}{llcccc}
\toprule
Type & Method & cost$(1,4)$ & cost$(1,9)$ & cost$(1,19)$ \\
\midrule 
\multirow{4}*{Individuals} & WSS & 8.5 &7.5 & 6.75 \\
~ & ZV  & 6 & 5.5 & 6 \\
~ & POIS  & 2.5 &4.25 & 4.5\\
~ & NB  &6.75 &6 & 6.25\\
\hline
\multirow{5}*{Combinations} & SA & 4.5& 5& 4.5 \\
~ & NSGA-\Romannum{3}-c  & 3.25& 3.75 & 4\\
~ & NSGA-\Romannum{3}-hs  & 5& 3.75 & 3.25\\
~ &  Cost-opt & 4.25 & 4.75 & 5\\
~ & DRPS-opt & 4.25 & 4.5& 4.75 \\
\bottomrule 
\end{tabular}}
\end{table}

Among individuals, POIS generally yields the superior DRPS, total cost, and holding cost but incurs high stockout cost, particularly in cost$(1,9)$ and cost$(1,19)$. Conversely, ZV performs best for  cost$(1,4)$ and cost$(1,9)$, while WSS excels at cost$(1,19)$. No single method dominates across all metrics, highlighting the inherent trade-offs in individual forecasting, especially under different service level regimes.

Single-objective combinations further magnify these trade-offs. Cost-opt minimizes cost and holding stock effectively but produces the worst DRPS. Conversely, DRPS-opt outperforms individual methods and Cost-opt in accuracy but performs poorly in holding stock. The SA consistently yields mediocre results, failing to significantly improve upon individual baselines.

\begin{figure}
    \centering
    \includegraphics[width=\linewidth]{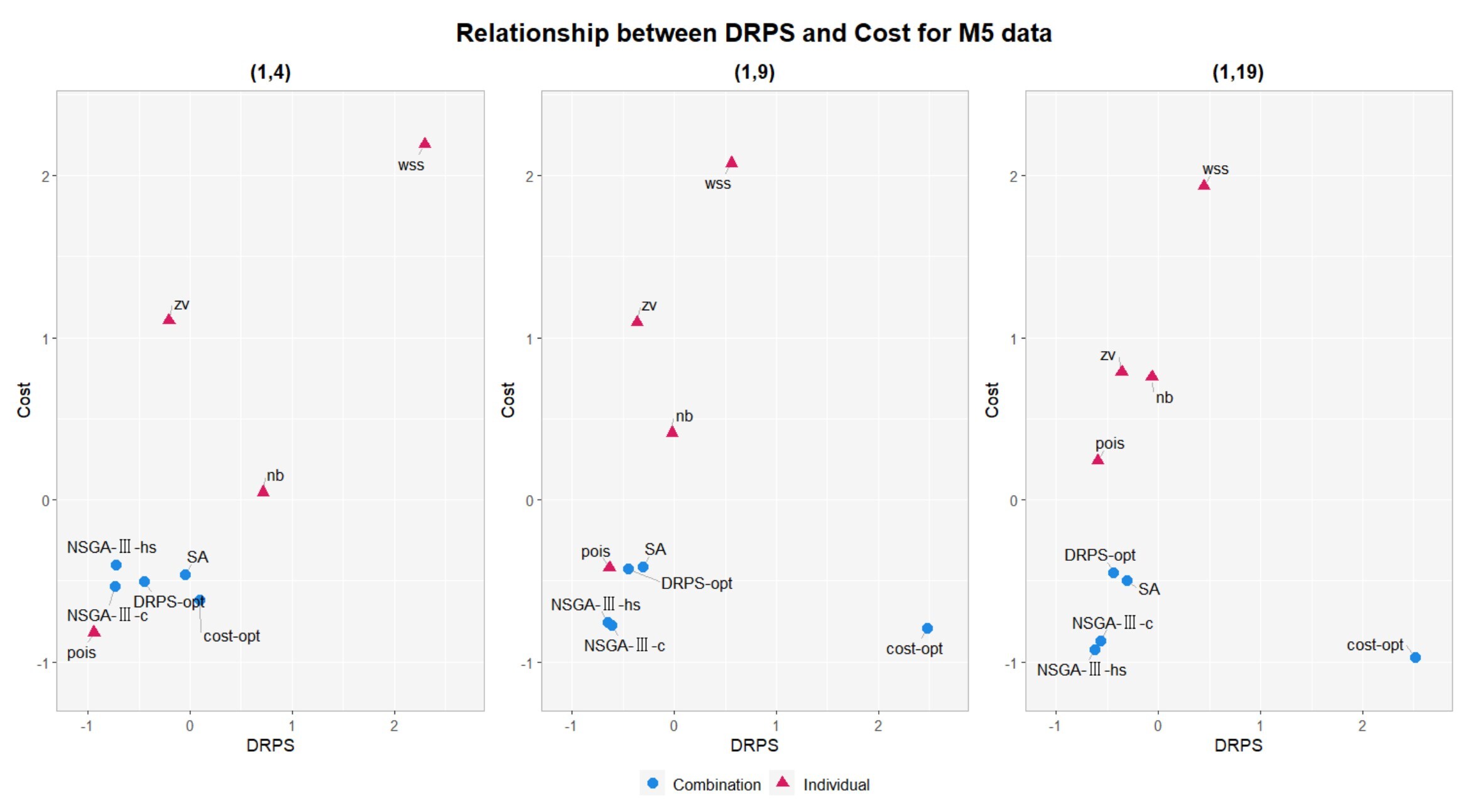}
    \caption{The relationship between DRPS and cost of M5 data. Each subplot represents a setting of unit cost. Metrics are $Z$-score normalized within each panel. }
    \label{fig:drps-cost}
\end{figure}

\begin{figure}
    \centering
    \includegraphics[width=\linewidth]{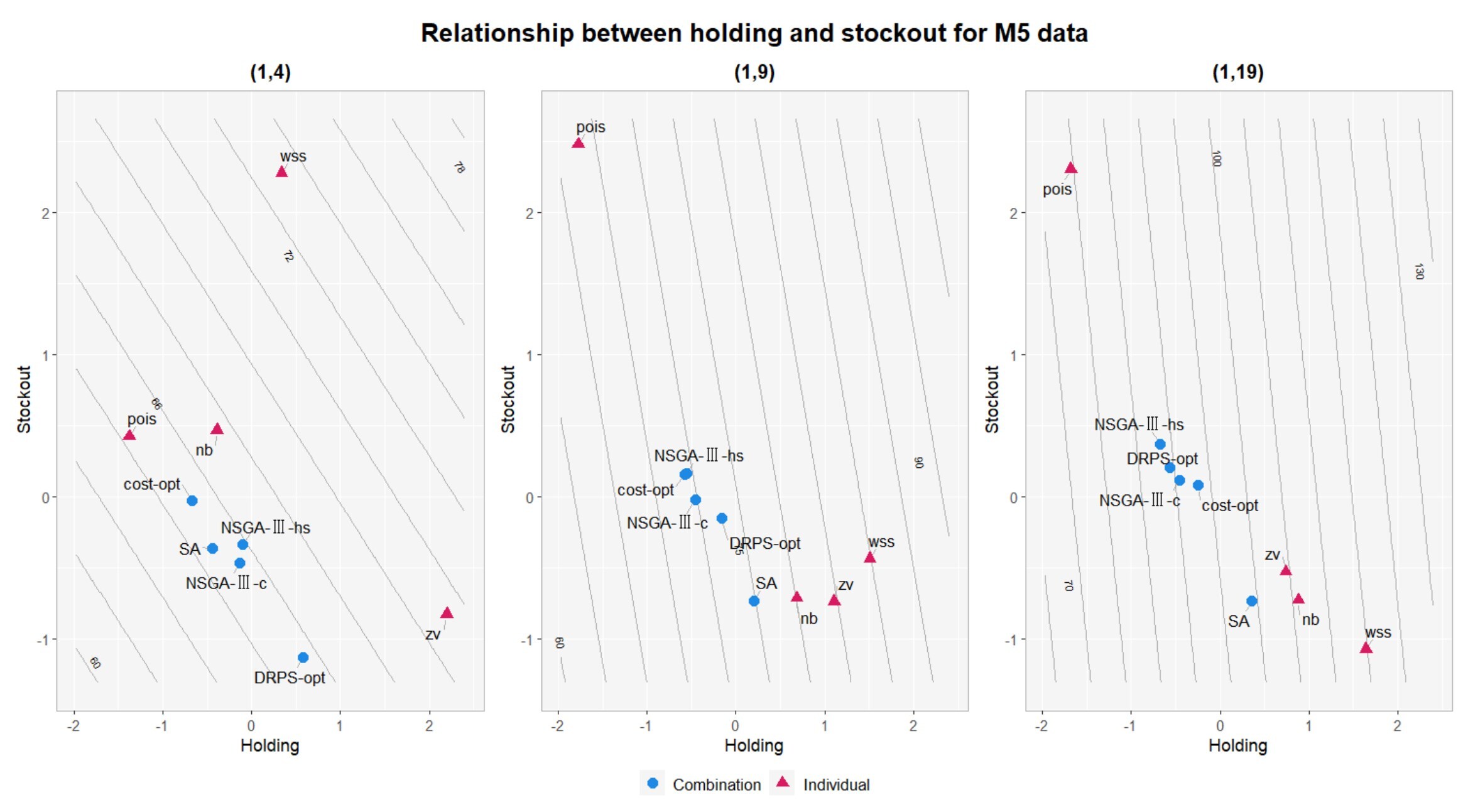}
    \caption{Trade-off between holding stock and stockout quantity for the M5 data under three cost scenarios, with total cost contours. Both axes are $Z$-score normalized within each panel, while contour lines represent total cost in the original scale.}
    \label{fig:holding-lost}
\end{figure}

By contrast, the MOO-based methods (NSGA-\Romannum{3}-c and NSGA-\Romannum{3}-hs) demonstrate superior robustness. While they rarely dominate all metrics simultaneously, they achieve the best average ranks (Table \ref{tab:single-rank}), indicating balanced performance across conflicting objectives. Figures \ref{fig:drps-cost} and \ref{fig:holding-lost} visualize these advantages. In Figure \ref{fig:drps-cost}, MOO solutions consistently occupy the efficient bottom-left region (minimizing both DRPS and Cost) across most scenarios. Similarly, Figure \ref{fig:holding-lost} shows that these methods achieve low levels of both holding stock and stockout quantity. They also balance the two under relatively low total cost, as indicated by the cost contours. These findings suggest that MOO-based methods can achieve both relatively accurate and economically efficient forecasts, while maintaining an appropriate balance across various decision targets under different conditions, rather than being biased toward a single performance dimension.

\subsection{Robustness to mismatch of unit cost}

Section \ref{m5-result} assumes that the decision-maker has perfect knowledge of the unit costs associated with holding stock and stockout quantity. In practice, however, such parameters are often subject to misspecification, and the perceived costs may differ from the true ones. To account for potential mismatches between the unit cost parameters used in optimization and those in real-world settings, we evaluate the performance of NSGA-\Romannum{3}-c and Cost-opt under small perturbations of the unit stockout cost $c_2$, which directly enters the optimization of both methods. Specifically, we consider proportional adjustments of $c_2$, namely $0.8c_2$, $0.9c_2$, $1.1c_2$ and $1.2c_2$.

The resulting cost evaluations under these perturbations are reported in Tables \ref{tab:robust-uncertainty-4}, \ref{tab:robust-uncertainty-9}, and \ref{tab:robust-uncertainty-19}. Compared with the baseline results in Tables \ref{tab:single-80}, \ref{tab:single-90}, and \ref{tab:single-95}, the relative rankings of the methods remain largely stable across different levels of perturbation. This suggests that moderate misspecification of cost parameters does not materially affect the cost performance of the decision-oriented forecast combinations, indicating that the proposed MOO framework exhibits robustness to uncertainty in cost specification. Consequently, the findings in Section \ref{m5-result} are empirically robust to moderate cost misestimation.

\begin{table}[!ht]
\centering
\caption{Cost under fluctuation for M5 data with cost parameters $(c_1,c_2)=(1,4)$. \textbf{Bold} numbers highlight the superior performance, while \textit{italics} ones indicate the second best results.}
\label{tab:robust-uncertainty-4}
\begin{tabular}{lccccc}
\toprule
 Method & 0.8$c_2$ & 0.9$c_2$ & 1.1$c_2$ & 1.2$c_2$ \\
\midrule 
 WSS & 65.7590 & 69.6441 & 77.4144& 81.3000 \\
 ZV  & 64.0892 & 67.1862 & 73.3801& 76.4770 \\
 POIS  & \textbf{57.7238} & \textbf{61.1385} & \textbf{67.9678}&  \textbf{71.3825}\\
 NB  &60.2727 & 63.6985 & 70.5502& 73.9760 \\
\hline
  SA & 59.2101& 62.4133 & 68.8195 & 72.0226\\
 NSGA-\Romannum{3}-c  & 59.0182& 62.2070& 68.5846&  71.7734\\
 NSGA-\Romannum{3}-hs  & 59.3512 & 62.5721 & 69.0139& 72.2348\\
  Cost-opt & \textit{58.6874} & \textit{61.9639} & 68.5171& 71.7936\\
 DRPS-opt & 59.4451 & 62.4652 & \textit{68.5053}& \textit{71.5254} \\
\bottomrule 
\end{tabular}
\end{table}

\begin{table}[!ht]
\centering
\caption{Cost under fluctuation for M5 data with cost parameters $(c_1,c_2)=(1,9)$. \textbf{Bold} numbers highlight the superior performance, while \textit{italics} ones indicate the second best results.}
\label{tab:robust-uncertainty-9}

\begin{tabular}{lccccc}
\toprule
 Method & 0.8$c_2$ & 0.9$c_2$ & 1.1$c_2$ & 1.2$c_2$ \\
\midrule 
 WSS & 97.7402 & 101.2225 & 108.1872& 111.6695 \\
 ZV  & 93.7618 & 97.1100 & 103.8063& 107.1545 \\
 POIS  & \textbf{84.3391} & 89.1186 & 98.6774&  103.4569\\
 NB  &90.7837& 94.1423 & 100.8594& 104.2180 \\
\hline
  SA & 87.2412& 90.5910 & 97.2905 & 100.6403\\
 NSGA-\Romannum{3}-c  & 85.0324 & 88.7000 & \textbf{96.0339}&  \textbf{99.7011}\\
 NSGA-\Romannum{3}-hs  & 85.2434 & 89.2285 & 97.2000& 101.1835\\
  Cost-opt & \textit{84.7980} & \textbf{88.5439} & \textit{96.0358}& \textit{99.7817}\\
 DRPS-opt & 90.4161 & 94.5791 & 102.9053& 107.0684 \\
\bottomrule 
\end{tabular}
\end{table}

\begin{table}[!ht]
\centering
\caption{Cost under fluctuation for M5 data with cost parameters $(c_1,c_2)=(1,19)$. \textbf{Bold} numbers highlight the superior performance, while \textit{italics} ones indicate the second best results.}
\label{tab:robust-uncertainty-19}

\begin{tabular}{lccccc}
\toprule
 Method & 0.8$c_2$ & 0.9$c_2$ & 1.1$c_2$ & 1.2$c_2$ \\
\midrule 
 WSS & 133.8050 & 136.7357 & 142.5971& 145.5278 \\
 ZV  & 125.4193 & 128.9750 & 136.0864& 139.6421 \\
 POIS  & 115.5422 & 122.3381 & 135.9299&  142.7258\\
 NB  &125.6910 & 129.0206 & 135.6797& 139.0093 \\
\hline
  SA & 117.9124& 121.2304 & 127.8665 & 131.1845\\
 NSGA-\Romannum{3}-c  & 113.6682 & 117.9613 & 126.5477&  \textit{130.8408}\\
 NSGA-\Romannum{3}-hs  & \textbf{112.7231} & \textit{117.3074} & \textit{126.4762}& 131.0606\\
  Cost-opt & \textit{112.8129} & \textbf{117.2096} & \textbf{126.0030}& \textbf{130.3997}\\
 DRPS-opt & 124.5882 & 130.0014 & 140.8276& 146.2407 \\
\bottomrule 
\end{tabular}
\end{table}

\section{Empirical Evaluation: RAF data}\label{case}

\subsection{Data description and experimental setup}

We evaluate our proposed framework using the Royal Air Force (RAF) aerospace spare parts dataset, a benchmark extensively utilized in inventory management and demand forecasting literature \citep{li2022bayesian, kourentzes2021elucidate, petropoulos2015forecast, teunter2009forecasting}. The data set includes 5,000 monthly time series, each with 84-period observations. 

As noted in \citet{li2022bayesian}, the RAF data exhibits higher intermittency than the M5 dataset, providing a rigorous test for the robustness of the four individual forecasting methods. We divide each series into three segments of lengths 72, 6, and 6 for obtaining base forecasts, weight estimation, and evaluation. In the absence of unit cost information, we adopt the same inventory policy and cost structures defined in Section \ref{empirical}, ensuring comparability across datasets.

\subsection{Performance results}

The performance metrics are presented in Tables \ref{tab:raf-result-4}, \ref{tab:raf-result-9}, and \ref{tab:raf-result-19}, while Table \ref{tab:single-rank-raf} summarizes the average ranks across the three cost scenarios.
 
A defining characteristic of the individual methods is the pronounced trade-off between competing metrics. Specifically, low holding costs typically coincide with high stockout quantity, and minimal DRPS rarely aligns with minimal total cost. Notably, in the high service-level scenarios (cost(1,9) and cost(1,19)), POIS achieves the lowest total cost despite yielding the highest (worst) DRPS. This discrepancy corroborates the findings of \citet{kourentzes2020optimising}, highlighting the limitations of selecting models based solely on forecast accuracy and reinforcing the inherent disconnect between statistical loss and decision value.

Regarding combinations, the MOO-based approach displays a clear capacity to balance these conflicting objectives. While MOO-based combinations rarely achieve the absolute best score in any single metric, they secure the best average ranks for cost(1,4)  and cost(1,9), and rank second only to DRPS-opt at cost(1,19). These results confirm the method's ability to effectively reconcile forecasting accuracy with decision-oriented goals, providing a stable compromise even in highly intermittent demand environments.

\begin{table}[!ht]
\centering
\caption{Forecast and decision metrics for RAF data with cost parameters $(c_1,c_2)=(1,4)$. \textbf{Bold} highlights the best performance, while \textit{italics} indicates the second best results.}
\label{tab:raf-result-4}
{\begin{tabular}{llccccc}
\toprule
Type & Method & DRPS & Cost & Holding & Stockout \\
\midrule 
\multirow{4}*{Individuals} & WSS & 1.8102 & \textbf{30.7704} & \textbf{0.2032} & 7.6418 \\
~ & ZV  & 1.7620 & 35.5802 & 6.9506 & 7.1574 \\
~ & POIS  & 2.0418 & 37.8284 & 11.2068 & \textbf{6.6554}\\
~ & NB  &1.7969 & \textit{31.5008} & \textit{1.2056} & 7.5738 \\
\hline
\multirow{5}*{Combinations} & SA & 1.6849 & 35.4914 & 7.7978 & 6.9234\\
~ & NSGA-\Romannum{3}-c  & 1.6215 & 33.9438 & 5.0518 & 7.2230\\
~ & NSGA-\Romannum{3}-hs  & \textit{1.5967} & 34.1272 & 5.3480 & 7.1948\\
~ &  Cost-opt & 1.7466 & 32.2214 & 2.2902 & 7.4828\\
~ & DRPS-opt & \textbf{1.5362} & 36.2018 & 8.7826 & \textit{6.8548}\\
\bottomrule 
\end{tabular}}
\end{table}

\begin{table}[!ht]
\centering
\caption{Forecast and decision metrics for RAF data with cost parameters $(c_1,c_2)=(1,9)$. \textbf{Bold} highlights the best performance, while \textit{italics} indicates the second best results.}
\label{tab:raf-result-9}
{\begin{tabular}{llccccc}
\toprule
Type & Method & DRPS & Cost & Holding & Stockout \\
\midrule 
\multirow{4}*{Individuals} & WSS & 1.8102 & 73.4756 & 17.4812 & 6.2216 \\
~ & ZV  & 1.7620 & 87.6900 & 44.0418 & \textbf{4.8498}  \\
~ & POIS  & 2.0418 & \textit{72.0842} & 14.5958 & 6.3876\\
~ & NB  &1.7969 & \textbf{70.9262} & \textbf{7.9028} & 7.0026 \\
\hline
\multirow{5}*{Combinations} & SA & 1.6849 & 73.1294 & 17.1314 & 6.2220\\
~ & NSGA-\Romannum{3}-c  & 1.6290 & 72.6534 & 16.3044 & 6.2610\\
~ & NSGA-\Romannum{3}-hs  & \textit{1.6206} & 72.5088 & \textit{14.3634} & 6.4606 \\
~ &  Cost-opt & 1.7636 & 72.4034 & 14.7404 & 6.4070\\
~ & DRPS-opt & \textbf{1.5362} & 73.8418 & 18.7708 & \textit{6.1190}\\
\bottomrule 
\end{tabular}}
\end{table}

\begin{table}[!ht]
\centering
\caption{Forecast and decision metrics for RAF data with cost parameters $(c_1,c_2)=(1,19)$. \textbf{Bold} highlights the best performance, while \textit{italics} indicates the second best results.}
\label{tab:raf-result-19}
{\begin{tabular}{llccccc}
\toprule
Type & Method & DRPS & Cost & Holding &Stockout \\
\midrule 
\multirow{4}*{Individuals} & WSS & 1.8102 & 137.5666 & 63.7554 & \textit{3.8848} \\
~ & ZV  & 1.7620 & 161.3476 & 104.4046 & \textbf{2.9970}  \\
~ & POIS  & 2.0418 & 135.0910 & \textbf{17.4582} & 6.1912\\
~ & NB  &1.7969 & 136.8174 & 39.1042 & 5.1428 \\
\hline
\multirow{5}*{Combinations} & SA & 1.6849 & \textit{132.6742} & 52.2548 & 4.2326 \\
~ & NSGA-\Romannum{3}-c  & 1.5774 & 134.3768 & 34.6382 & 5.2494\\
~ & NSGA-\Romannum{3}-hs  & \textit{1.5712} & 134.1996 & \textit{30.4900} & 5.4584 \\
~ &  Cost-opt & 1.8098 & 135.8908 & 34.1306 & 5.3558\\
~ & DRPS-opt & \textbf{1.5362} & \textbf{132.5148} & 39.9088 & 4.8740\\
\bottomrule 
\end{tabular}}
\end{table}

\begin{table}[!ht]
\centering
\caption{Average ranks of forecast and decision performance in three cost scenarios for RAF data.}
\label{tab:single-rank-raf}
{\begin{tabular}{llcccc}
\toprule
Type & Method & cost$(1,4)$ & cost$(1,9)$ & cost$(1,19)$ \\
\midrule 
\multirow{4}*{Individuals} & WSS & 4.75 & 6.25 & 6.5  \\
~ & ZV  & 5.75 & 6 & 6  \\
~ & POIS  & 7 & 5 & 6\\
~ & NB  &4.75 & 4.5 & 5.75\\
\hline
\multirow{5}*{Combinations} & SA & 5 & 5 & 4 \\
~ & NSGA-\Romannum{3}-c  & 4.25 & 4.5 & 4.25\\
~ & NSGA-\Romannum{3}-hs  & 4.25 & 4 & 3.75\\
~ &  Cost-opt & 4.5 & 5 & 5.75\\
~ & DRPS-opt & 4.75 & 4.75 & 3  \\
\bottomrule 
\end{tabular}}
\end{table}

\begin{figure}
    \centering
    \includegraphics[width=\linewidth]{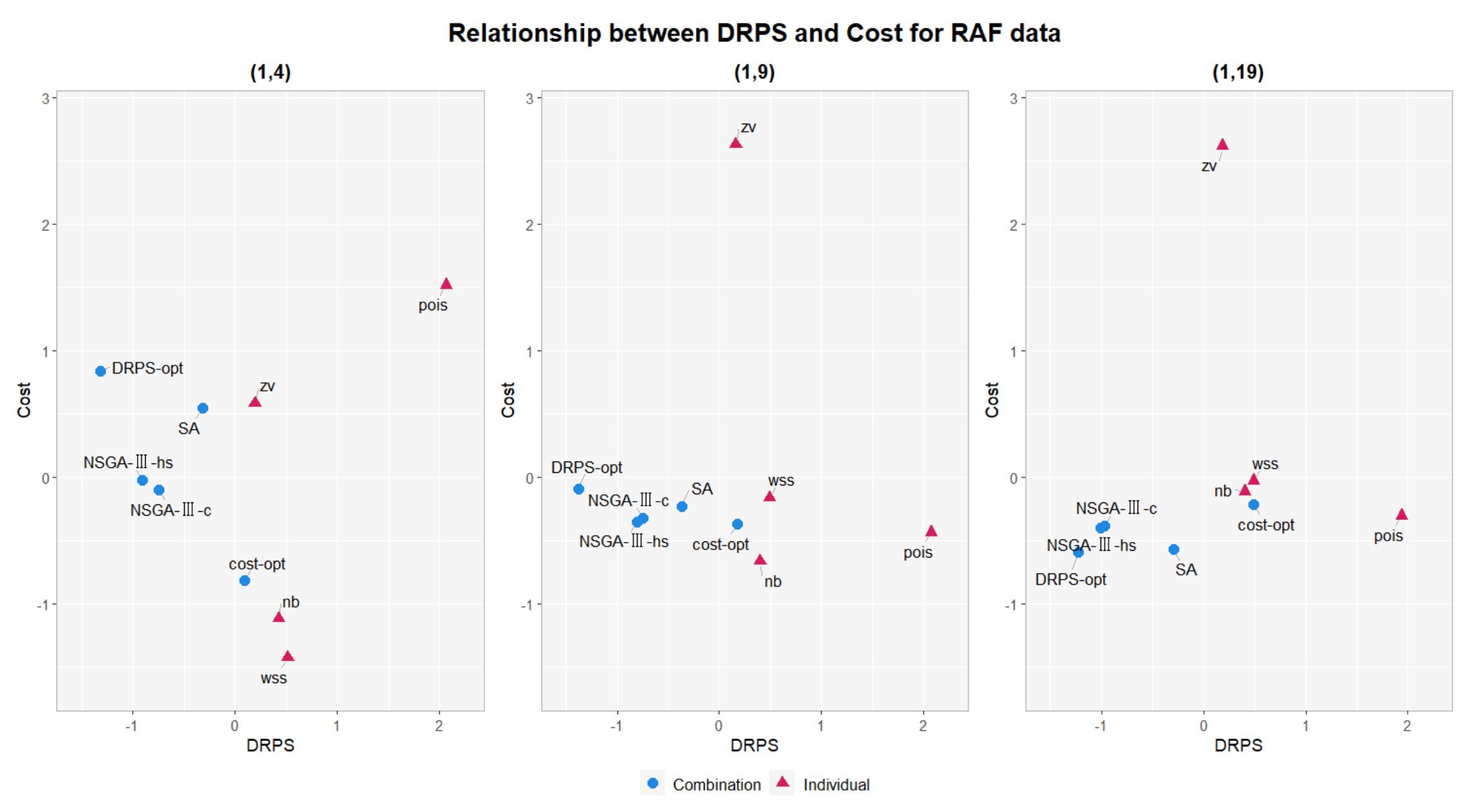}
    \caption{The relationship between DRPS and cost for the RAF data. The settings are similar to those in Figure \ref{fig:drps-cost}.}
    \label{fig:drps-cost-raf}
\end{figure}

\begin{figure}
    \centering
    \includegraphics[width=\linewidth]{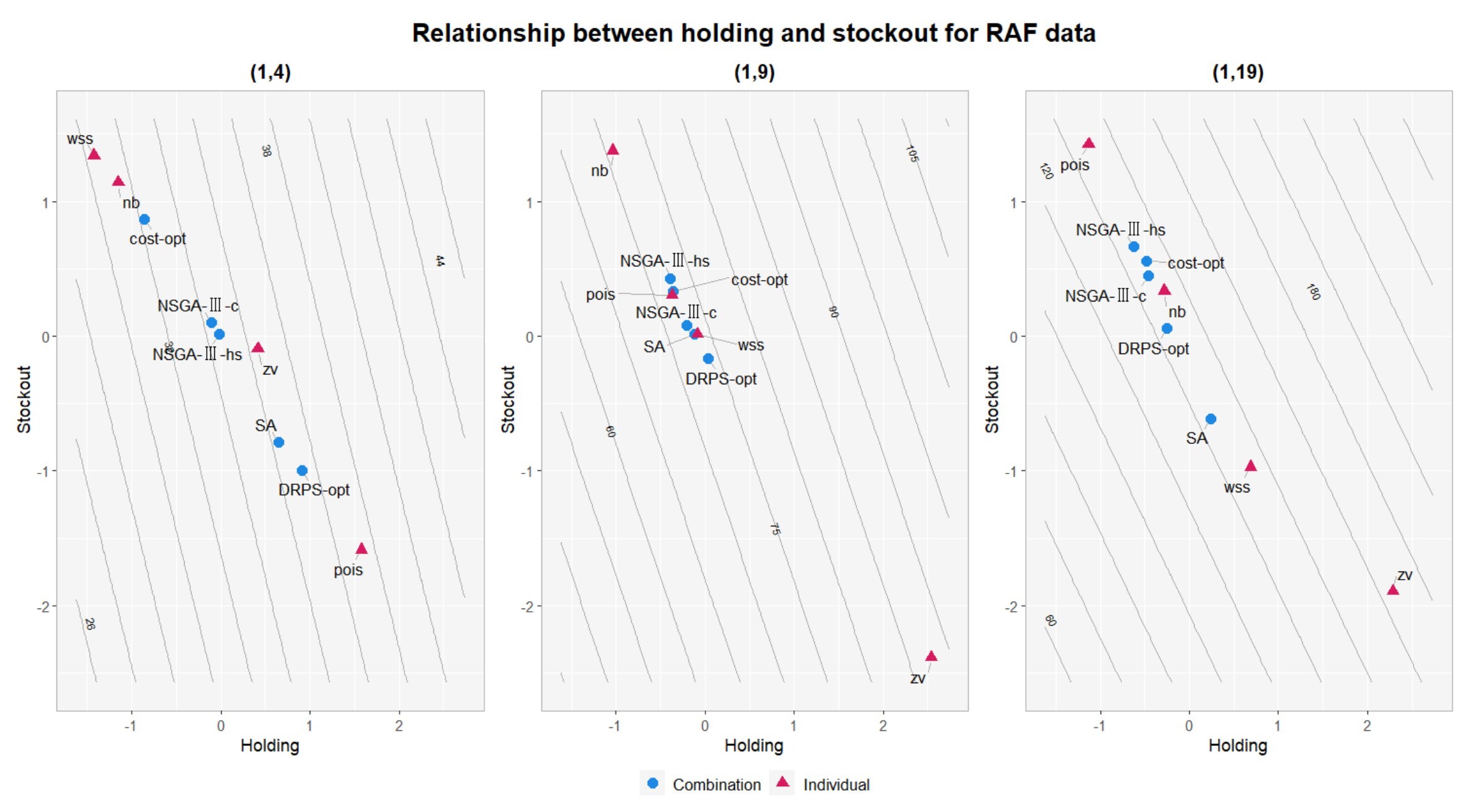}
    \caption{Trade-off between holding stock and stockout quantity for the RAF data, with total cost contours. Settings are consistent with those in Figure \ref{fig:holding-lost}.}
    \label{fig:hold-cost-raf}
\end{figure}

To visualize the trade-offs between conflicting objectives, Figure \ref{fig:drps-cost-raf} plots DRPS against total cost, while Figure \ref{fig:hold-cost-raf} shows the relationship between holding stock and stockout quantity. Figure \ref{fig:drps-cost-raf} reveals a clear separation between individual forecasting methods and combination strategies. Notably, all combination methods cluster in the bottom-left region—the ideal zone for minimizing both metrics simultaneously. Among these, the MOO combinations lie closest to the origin in the first two cost scenarios, indicating superior Pareto efficiency. Similarly, Figure \ref{fig:hold-cost-raf} shows that the MOO combinations occupy a central position along the efficiency frontier, rather than being concentrated at either extreme. By avoiding the extremes of excessive holding stock or frequent stockouts, these methods effectively balance operational risks to minimize total cost.

These visual results corroborate our findings from the M5 dataset, highlighting the ability of MOO-based combinations to resolve the inherent conflicts between forecasting accuracy and decision quality. By consistently achieving favorable trade-offs across different domains and demand patterns, these findings support the generalizability and robustness of the proposed framework.

\section{Conclusion}\label{conclusion}

This paper proposes a novel framework for probabilistic forecast combination based on MOO, designed to bridge the gap between statistical forecasting and inventory decision-making. Our approach introduces a forecast-decision alignment mechanism that integrates demand forecasting with inventory control by simultaneously balancing conflicting objectives. A key advantage of this framework is its flexibility: it is model-agnostic and does not impose restrictions on the underlying component models. This allows practitioners to adapt general-purpose forecasting models to specific decision environments without requiring complex, task-specific modifications to their native loss functions, thereby preserving modularity between forecasting and optimization components.

Empirical evaluations using the Walmart and RAF spare parts datasets demonstrate that MOO-based forecast combinations achieve a superior trade-off between forecasting accuracy and decision quality compared to traditional methods. Notably, our results indicate that these objectives are not always competing; in several instances, the proposed method improved both forecast accuracy and inventory performance simultaneously, outperforming individual models, simple averaging, and single-objective optimization combinations, highlighting the potential complementarity between statistical and operational objectives.

This research highlights two critical insights for the field of integrated forecasting and inventory management. First, simultaneously optimizing for forecast errors and decision targets is highly effective. Through MOO, distinct goals can be harmonized, revealing synergies that single-objective approaches often overlook. Second, probabilistic forecast combination proves to be an efficient strategy for adapting statistical forecasting methods to operational decision tasks. It provides a practical alternative to retraining complex models from scratch, particularly in settings where data sparsity or model heterogeneity makes end-to-end training impractical.

Despite its advantages, the proposed framework has some limitations. First, the use of NSGA-III introduces non-negligible computational overhead, particularly in large-scale or real-time applications. Second, the framework relies on a fixed validation window for weight estimation, which may be sensitive to structural changes in demand patterns. Third, the inventory setting is restricted to a single-period, single-item newsvendor structure, limiting direct applicability to more complex multi-period or multi-echelon systems.

Future research directions include investigating alternative MOO algorithms to further enhance computational efficiency and solution quality. Additionally, there is potential to extend this philosophy beyond combination methods, developing end-to-end approaches that modify the training of traditional forecasting models to align directly with decision-making objectives, as well as exploring dynamic (time-varying) Pareto solutions in non-stationary demand environments.

\section*{Disclosure statement}

The authors report there are no competing interests to declare.

\section*{Data availability statement}

The M5 competition \citep{makridakis2022m5} dataset is available at \url{https://github.com/Mcompetitions/M5-methods}. The RAF dataset has been used in previous literature \citep{teunter2009forecasting,petropoulos2015forecast, kourentzes2021elucidate,li2022bayesian} and is available upon request.

\section*{Funding}

This work was supported by the the National Natural Science Foundation of China [Grant numbers: 72571014, 72171011].




\bibliographystyle{apacite}
\bibliography{references}

\begin{thebibliography}{}

\bibitem [\protect \citeauthoryear {%
Baran%
\ \BBA {} Lerch%
}{%
Baran%
\ \BBA {} Lerch%
}{%
{\protect \APACyear {2018}}%
}]{%
baran2018combining}
\APACinsertmetastar {%
baran2018combining}%
\begin{APACrefauthors}%
Baran, S.%
\BCBT {}\ \BBA {} Lerch, S.%
\end{APACrefauthors}%
\unskip\
\newblock
\APACrefYearMonthDay{2018}{}{}.
\newblock
{\BBOQ}\APACrefatitle {Combining predictive distributions for the statistical post-processing of ensemble forecasts} {Combining predictive distributions for the statistical post-processing of ensemble forecasts}.{\BBCQ}
\newblock
\APACjournalVolNumPages{International Journal of Forecasting}{34}{3}{477--496}.
\newblock
\begin{APACrefDOI} \doi{10.1016/j.ijforecast.2018.01.005} \end{APACrefDOI}
\PrintBackRefs{\CurrentBib}

\bibitem [\protect \citeauthoryear {%
Bates%
\ \BBA {} Granger%
}{%
Bates%
\ \BBA {} Granger%
}{%
{\protect \APACyear {1969}}%
}]{%
bates1969combination}
\APACinsertmetastar {%
bates1969combination}%
\begin{APACrefauthors}%
Bates, J\BPBI M.%
\BCBT {}\ \BBA {} Granger, C\BPBI W.%
\end{APACrefauthors}%
\unskip\
\newblock
\APACrefYearMonthDay{1969}{}{}.
\newblock
{\BBOQ}\APACrefatitle {The combination of forecasts} {The combination of forecasts}.{\BBCQ}
\newblock
\APACjournalVolNumPages{Journal of the Operational Research Society}{20}{4}{451--468}.
\newblock
\begin{APACrefDOI} \doi{10.1057/jors.1969.103} \end{APACrefDOI}
\PrintBackRefs{\CurrentBib}

\bibitem [\protect \citeauthoryear {%
Bertsimas%
, McCord%
\BCBL {}\ \BBA {} Sturt%
}{%
Bertsimas%
\ \protect \BOthers {.}}{%
{\protect \APACyear {2023}}%
}]{%
bertsimas2023dynamic}
\APACinsertmetastar {%
bertsimas2023dynamic}%
\begin{APACrefauthors}%
Bertsimas, D.%
, McCord, C.%
\BCBL {}\ \BBA {} Sturt, B.%
\end{APACrefauthors}%
\unskip\
\newblock
\APACrefYearMonthDay{2023}{}{}.
\newblock
{\BBOQ}\APACrefatitle {Dynamic optimization with side information} {Dynamic optimization with side information}.{\BBCQ}
\newblock
\APACjournalVolNumPages{European Journal of Operational Research}{304}{2}{634--651}.
\newblock
\begin{APACrefDOI} \doi{10.1016/j.ejor.2022.03.030} \end{APACrefDOI}
\PrintBackRefs{\CurrentBib}

\bibitem [\protect \citeauthoryear {%
{Blank}%
\ \BBA {} {Deb}%
}{%
{Blank}%
\ \BBA {} {Deb}%
}{%
{\protect \APACyear {2020}}%
}]{%
pymoo}
\APACinsertmetastar {%
pymoo}%
\begin{APACrefauthors}%
{Blank}, J.%
\BCBT {}\ \BBA {} {Deb}, K.%
\end{APACrefauthors}%
\unskip\
\newblock
\APACrefYearMonthDay{2020}{}{}.
\newblock
{\BBOQ}\APACrefatitle {{Pymoo: Multi-Objective Optimization in Python}} {{Pymoo: Multi-Objective Optimization in Python}}.{\BBCQ}
\newblock
\APACjournalVolNumPages{IEEE Access}{8}{}{89497-89509}.
\newblock
\begin{APACrefDOI} \doi{10.1109/ACCESS.2020.2990567} \end{APACrefDOI}
\PrintBackRefs{\CurrentBib}

\bibitem [\protect \citeauthoryear {%
Chen%
, Jin%
, Zhou%
\BCBL {}\ \BBA {} Tian%
}{%
Chen%
\ \protect \BOthers {.}}{%
{\protect \APACyear {2025}}%
}]{%
chen2025novel}
\APACinsertmetastar {%
chen2025novel}%
\begin{APACrefauthors}%
Chen, Y.%
, Jin, M.%
, Zhou, Z.%
\BCBL {}\ \BBA {} Tian, Z.%
\end{APACrefauthors}%
\unskip\
\newblock
\APACrefYearMonthDay{2025}{}{}.
\newblock
{\BBOQ}\APACrefatitle {A novel ensemble learning framework based on news sentiment enhancement and multi-objective optimizer for carbon price forecasting} {A novel ensemble learning framework based on news sentiment enhancement and multi-objective optimizer for carbon price forecasting}.{\BBCQ}
\newblock
\APACjournalVolNumPages{Computational Economics}{66}{5}{3709--3733}.
\newblock
\begin{APACrefDOI} \doi{10.1007/s10614-024-10828-6} \end{APACrefDOI}
\PrintBackRefs{\CurrentBib}

\bibitem [\protect \citeauthoryear {%
Conflitti%
, De~Mol%
\BCBL {}\ \BBA {} Giannone%
}{%
Conflitti%
\ \protect \BOthers {.}}{%
{\protect \APACyear {2015}}%
}]{%
conflitti2015optimal}
\APACinsertmetastar {%
conflitti2015optimal}%
\begin{APACrefauthors}%
Conflitti, C.%
, De~Mol, C.%
\BCBL {}\ \BBA {} Giannone, D.%
\end{APACrefauthors}%
\unskip\
\newblock
\APACrefYearMonthDay{2015}{}{}.
\newblock
{\BBOQ}\APACrefatitle {Optimal combination of survey forecasts} {Optimal combination of survey forecasts}.{\BBCQ}
\newblock
\APACjournalVolNumPages{International Journal of Forecasting}{31}{4}{1096--1103}.
\newblock
\begin{APACrefDOI} \doi{10.1016/j.ijforecast.2015.03.009} \end{APACrefDOI}
\PrintBackRefs{\CurrentBib}

\bibitem [\protect \citeauthoryear {%
Deb%
\ \BBA {} Jain%
}{%
Deb%
\ \BBA {} Jain%
}{%
{\protect \APACyear {2014}}%
}]{%
deb2013evolutionary}
\APACinsertmetastar {%
deb2013evolutionary}%
\begin{APACrefauthors}%
Deb, K.%
\BCBT {}\ \BBA {} Jain, H.%
\end{APACrefauthors}%
\unskip\
\newblock
\APACrefYearMonthDay{2014}{}{}.
\newblock
{\BBOQ}\APACrefatitle {An evolutionary many-objective optimization algorithm using reference-point-based nondominated sorting approach, part I: solving problems with box constraints} {An evolutionary many-objective optimization algorithm using reference-point-based nondominated sorting approach, part i: solving problems with box constraints}.{\BBCQ}
\newblock
\APACjournalVolNumPages{IEEE Transactions on Evolutionary Computation}{18}{4}{577--601}.
\newblock
\begin{APACrefDOI} \doi{10.1109/TEVC.2013.2281535} \end{APACrefDOI}
\PrintBackRefs{\CurrentBib}

\bibitem [\protect \citeauthoryear {%
Del~Negro%
, Hasegawa%
\BCBL {}\ \BBA {} Schorfheide%
}{%
Del~Negro%
\ \protect \BOthers {.}}{%
{\protect \APACyear {2016}}%
}]{%
del2016dynamic}
\APACinsertmetastar {%
del2016dynamic}%
\begin{APACrefauthors}%
Del~Negro, M.%
, Hasegawa, R\BPBI B.%
\BCBL {}\ \BBA {} Schorfheide, F.%
\end{APACrefauthors}%
\unskip\
\newblock
\APACrefYearMonthDay{2016}{}{}.
\newblock
{\BBOQ}\APACrefatitle {{Dynamic prediction pools: An investigation of financial frictions and forecasting performance}} {{Dynamic prediction pools: An investigation of financial frictions and forecasting performance}}.{\BBCQ}
\newblock
\APACjournalVolNumPages{Journal of Econometrics}{192}{2}{391--405}.
\newblock
\begin{APACrefDOI} \doi{10.1016/j.jeconom.2016.02.006} \end{APACrefDOI}
\PrintBackRefs{\CurrentBib}

\bibitem [\protect \citeauthoryear {%
Diebold%
, Shin%
\BCBL {}\ \BBA {} Zhang%
}{%
Diebold%
\ \protect \BOthers {.}}{%
{\protect \APACyear {2023}}%
}]{%
diebold2023aggregation}
\APACinsertmetastar {%
diebold2023aggregation}%
\begin{APACrefauthors}%
Diebold, F\BPBI X.%
, Shin, M.%
\BCBL {}\ \BBA {} Zhang, B.%
\end{APACrefauthors}%
\unskip\
\newblock
\APACrefYearMonthDay{2023}{}{}.
\newblock
{\BBOQ}\APACrefatitle {{On the aggregation of probability assessments: Regularized mixtures of predictive densities for Eurozone inflation and real interest rates}} {{On the aggregation of probability assessments: Regularized mixtures of predictive densities for Eurozone inflation and real interest rates}}.{\BBCQ}
\newblock
\APACjournalVolNumPages{Journal of Econometrics}{237}{2}{105321}.
\newblock
\begin{APACrefDOI} \doi{10.1016/j.jeconom.2022.06.008} \end{APACrefDOI}
\PrintBackRefs{\CurrentBib}

\bibitem [\protect \citeauthoryear {%
Donti%
, Amos%
\BCBL {}\ \BBA {} Kolter%
}{%
Donti%
\ \protect \BOthers {.}}{%
{\protect \APACyear {2017}}%
}]{%
NIPS2017_3fc2c60b}
\APACinsertmetastar {%
NIPS2017_3fc2c60b}%
\begin{APACrefauthors}%
Donti, P.%
, Amos, B.%
\BCBL {}\ \BBA {} Kolter, J\BPBI Z.%
\end{APACrefauthors}%
\unskip\
\newblock
\APACrefYearMonthDay{2017}{}{}.
\newblock
{\BBOQ}\APACrefatitle {Task-based end-to-end model learning in stochastic optimization} {Task-based end-to-end model learning in stochastic optimization}.{\BBCQ}
\newblock
\APACjournalVolNumPages{Advances in Neural Information Processing Systems}{30}{}{5484-5494}.
\PrintBackRefs{\CurrentBib}

\bibitem [\protect \citeauthoryear {%
Elmachtoub%
\ \BBA {} Grigas%
}{%
Elmachtoub%
\ \BBA {} Grigas%
}{%
{\protect \APACyear {2022}}%
}]{%
elmachtoub2022smart}
\APACinsertmetastar {%
elmachtoub2022smart}%
\begin{APACrefauthors}%
Elmachtoub, A\BPBI N.%
\BCBT {}\ \BBA {} Grigas, P.%
\end{APACrefauthors}%
\unskip\
\newblock
\APACrefYearMonthDay{2022}{}{}.
\newblock
{\BBOQ}\APACrefatitle {Smart “predict, then optimize”} {Smart “predict, then optimize”}.{\BBCQ}
\newblock
\APACjournalVolNumPages{Management Science}{68}{1}{9--26}.
\newblock
\begin{APACrefDOI} \doi{10.1287/mnsc.2020.3922} \end{APACrefDOI}
\PrintBackRefs{\CurrentBib}

\bibitem [\protect \citeauthoryear {%
Ferreira%
, Lee%
\BCBL {}\ \BBA {} Simchi-Levi%
}{%
Ferreira%
\ \protect \BOthers {.}}{%
{\protect \APACyear {2016}}%
}]{%
ferreira2016analytics}
\APACinsertmetastar {%
ferreira2016analytics}%
\begin{APACrefauthors}%
Ferreira, K\BPBI J.%
, Lee, B\BPBI H\BPBI A.%
\BCBL {}\ \BBA {} Simchi-Levi, D.%
\end{APACrefauthors}%
\unskip\
\newblock
\APACrefYearMonthDay{2016}{}{}.
\newblock
{\BBOQ}\APACrefatitle {{Analytics for an online retailer: Demand forecasting and price optimization}} {{Analytics for an online retailer: Demand forecasting and price optimization}}.{\BBCQ}
\newblock
\APACjournalVolNumPages{Manufacturing \& Service Operations Management}{18}{1}{69--88}.
\newblock
\begin{APACrefDOI} \doi{10.1287/msom.2015.0561} \end{APACrefDOI}
\PrintBackRefs{\CurrentBib}

\bibitem [\protect \citeauthoryear {%
Ganjehkaviri%
, Jaafar%
, Hosseini%
\BCBL {}\ \BBA {} Barzegaravval%
}{%
Ganjehkaviri%
\ \protect \BOthers {.}}{%
{\protect \APACyear {2017}}%
}]{%
ganjehkaviri2017genetic}
\APACinsertmetastar {%
ganjehkaviri2017genetic}%
\begin{APACrefauthors}%
Ganjehkaviri, A.%
, Jaafar, M\BPBI M.%
, Hosseini, S\BPBI E.%
\BCBL {}\ \BBA {} Barzegaravval, H.%
\end{APACrefauthors}%
\unskip\
\newblock
\APACrefYearMonthDay{2017}{}{}.
\newblock
{\BBOQ}\APACrefatitle {{Genetic algorithm for optimization of energy systems: Solution uniqueness, accuracy, Pareto convergence and dimension reduction}} {{Genetic algorithm for optimization of energy systems: Solution uniqueness, accuracy, Pareto convergence and dimension reduction}}.{\BBCQ}
\newblock
\APACjournalVolNumPages{Energy}{119}{}{167--177}.
\newblock
\begin{APACrefDOI} \doi{10.1016/j.energy.2016.12.034} \end{APACrefDOI}
\PrintBackRefs{\CurrentBib}

\bibitem [\protect \citeauthoryear {%
Garratt%
, Lee%
, Pesaran%
\BCBL {}\ \BBA {} Shin%
}{%
Garratt%
\ \protect \BOthers {.}}{%
{\protect \APACyear {2003}}%
}]{%
garratt2003forecast}
\APACinsertmetastar {%
garratt2003forecast}%
\begin{APACrefauthors}%
Garratt, A.%
, Lee, K.%
, Pesaran, M\BPBI H.%
\BCBL {}\ \BBA {} Shin, Y.%
\end{APACrefauthors}%
\unskip\
\newblock
\APACrefYearMonthDay{2003}{}{}.
\newblock
{\BBOQ}\APACrefatitle {{Forecast uncertainties in macroeconomic modeling: An application to the UK economy}} {{Forecast uncertainties in macroeconomic modeling: An application to the UK economy}}.{\BBCQ}
\newblock
\APACjournalVolNumPages{Journal of the American Statistical Association}{98}{464}{829--838}.
\newblock
\begin{APACrefDOI} \doi{10.1198/016214503000000765} \end{APACrefDOI}
\PrintBackRefs{\CurrentBib}

\bibitem [\protect \citeauthoryear {%
Geweke%
\ \BBA {} Amisano%
}{%
Geweke%
\ \BBA {} Amisano%
}{%
{\protect \APACyear {2011}}%
}]{%
geweke2011optimal}
\APACinsertmetastar {%
geweke2011optimal}%
\begin{APACrefauthors}%
Geweke, J.%
\BCBT {}\ \BBA {} Amisano, G.%
\end{APACrefauthors}%
\unskip\
\newblock
\APACrefYearMonthDay{2011}{}{}.
\newblock
{\BBOQ}\APACrefatitle {Optimal prediction pools} {Optimal prediction pools}.{\BBCQ}
\newblock
\APACjournalVolNumPages{Journal of Econometrics}{164}{1}{130--141}.
\newblock
\begin{APACrefDOI} \doi{10.1016/j.jeconom.2011.02.017} \end{APACrefDOI}
\PrintBackRefs{\CurrentBib}

\bibitem [\protect \citeauthoryear {%
Gneiting%
\ \BBA {} Ranjan%
}{%
Gneiting%
\ \BBA {} Ranjan%
}{%
{\protect \APACyear {2013}}%
}]{%
gneiting2013combining}
\APACinsertmetastar {%
gneiting2013combining}%
\begin{APACrefauthors}%
Gneiting, T.%
\BCBT {}\ \BBA {} Ranjan, R.%
\end{APACrefauthors}%
\unskip\
\newblock
\APACrefYearMonthDay{2013}{}{}.
\newblock
{\BBOQ}\APACrefatitle {Combining predictive distributions} {Combining predictive distributions}.{\BBCQ}
\newblock
\APACjournalVolNumPages{Electronic Journal of Statistics}{7}{}{1747--1782}.
\newblock
\begin{APACrefDOI} \doi{10.1214/13-EJS823} \end{APACrefDOI}
\PrintBackRefs{\CurrentBib}

\bibitem [\protect \citeauthoryear {%
Goltsos%
, Syntetos%
, Glock%
\BCBL {}\ \BBA {} Ioannou%
}{%
Goltsos%
\ \protect \BOthers {.}}{%
{\protect \APACyear {2022}}%
}]{%
goltsos2022inventory}
\APACinsertmetastar {%
goltsos2022inventory}%
\begin{APACrefauthors}%
Goltsos, T\BPBI E.%
, Syntetos, A\BPBI A.%
, Glock, C\BPBI H.%
\BCBL {}\ \BBA {} Ioannou, G.%
\end{APACrefauthors}%
\unskip\
\newblock
\APACrefYearMonthDay{2022}{}{}.
\newblock
{\BBOQ}\APACrefatitle {{Inventory--forecasting: Mind the gap}} {{Inventory--forecasting: Mind the gap}}.{\BBCQ}
\newblock
\APACjournalVolNumPages{European Journal of Operational Research}{299}{2}{397--419}.
\newblock
\begin{APACrefDOI} \doi{10.1016/j.ejor.2021.07.040} \end{APACrefDOI}
\PrintBackRefs{\CurrentBib}

\bibitem [\protect \citeauthoryear {%
Hall%
\ \BBA {} Mitchell%
}{%
Hall%
\ \BBA {} Mitchell%
}{%
{\protect \APACyear {2007}}%
}]{%
hall2007combining}
\APACinsertmetastar {%
hall2007combining}%
\begin{APACrefauthors}%
Hall, S\BPBI G.%
\BCBT {}\ \BBA {} Mitchell, J.%
\end{APACrefauthors}%
\unskip\
\newblock
\APACrefYearMonthDay{2007}{}{}.
\newblock
{\BBOQ}\APACrefatitle {Combining density forecasts} {Combining density forecasts}.{\BBCQ}
\newblock
\APACjournalVolNumPages{International Journal of Forecasting}{23}{1}{1--13}.
\newblock
\begin{APACrefDOI} \doi{10.1016/j.ijforecast.2006.08.001} \end{APACrefDOI}
\PrintBackRefs{\CurrentBib}

\bibitem [\protect \citeauthoryear {%
Hao%
, Zhao%
, Zhang%
, Cao%
\BCBL {}\ \BBA {} Li%
}{%
Hao%
\ \protect \BOthers {.}}{%
{\protect \APACyear {2024}}%
}]{%
hao2024constrained}
\APACinsertmetastar {%
hao2024constrained}%
\begin{APACrefauthors}%
Hao, Y.%
, Zhao, C.%
, Zhang, Y.%
, Cao, Y.%
\BCBL {}\ \BBA {} Li, Z.%
\end{APACrefauthors}%
\unskip\
\newblock
\APACrefYearMonthDay{2024}{}{}.
\newblock
{\BBOQ}\APACrefatitle {Constrained multi-objective optimization problems: Methodologies, algorithms and applications} {Constrained multi-objective optimization problems: Methodologies, algorithms and applications}.{\BBCQ}
\newblock
\APACjournalVolNumPages{Knowledge-Based Systems}{299}{}{111998}.
\newblock
\begin{APACrefDOI} \doi{10.1016/j.knosys.2024.111998} \end{APACrefDOI}
\PrintBackRefs{\CurrentBib}

\bibitem [\protect \citeauthoryear {%
Hora%
}{%
Hora%
}{%
{\protect \APACyear {2004}}%
}]{%
hora2004probability}
\APACinsertmetastar {%
hora2004probability}%
\begin{APACrefauthors}%
Hora, S\BPBI C.%
\end{APACrefauthors}%
\unskip\
\newblock
\APACrefYearMonthDay{2004}{}{}.
\newblock
{\BBOQ}\APACrefatitle {{Probability judgments for continuous quantities: Linear combinations and calibration}} {{Probability judgments for continuous quantities: Linear combinations and calibration}}.{\BBCQ}
\newblock
\APACjournalVolNumPages{Management Science}{50}{5}{597--604}.
\newblock
\begin{APACrefDOI} \doi{10.1287/mnsc.1040.0205} \end{APACrefDOI}
\PrintBackRefs{\CurrentBib}

\bibitem [\protect \citeauthoryear {%
Huber%
, M{\"u}ller%
, Fleischmann%
\BCBL {}\ \BBA {} Stuckenschmidt%
}{%
Huber%
\ \protect \BOthers {.}}{%
{\protect \APACyear {2019}}%
}]{%
huber2019data}
\APACinsertmetastar {%
huber2019data}%
\begin{APACrefauthors}%
Huber, J.%
, M{\"u}ller, S.%
, Fleischmann, M.%
\BCBL {}\ \BBA {} Stuckenschmidt, H.%
\end{APACrefauthors}%
\unskip\
\newblock
\APACrefYearMonthDay{2019}{}{}.
\newblock
{\BBOQ}\APACrefatitle {{A data-driven newsvendor problem: From data to decision}} {{A data-driven newsvendor problem: From data to decision}}.{\BBCQ}
\newblock
\APACjournalVolNumPages{European Journal of Operational Research}{278}{3}{904--915}.
\newblock
\begin{APACrefDOI} \doi{10.1016/j.ejor.2019.04.043} \end{APACrefDOI}
\PrintBackRefs{\CurrentBib}

\bibitem [\protect \citeauthoryear {%
Jain%
\ \BBA {} Deb%
}{%
Jain%
\ \BBA {} Deb%
}{%
{\protect \APACyear {2014}}%
}]{%
jain2013evolutionary}
\APACinsertmetastar {%
jain2013evolutionary}%
\begin{APACrefauthors}%
Jain, H.%
\BCBT {}\ \BBA {} Deb, K.%
\end{APACrefauthors}%
\unskip\
\newblock
\APACrefYearMonthDay{2014}{}{}.
\newblock
{\BBOQ}\APACrefatitle {An evolutionary many-objective optimization algorithm using reference-point based nondominated sorting approach, part II: Handling constraints and extending to an adaptive approach} {An evolutionary many-objective optimization algorithm using reference-point based nondominated sorting approach, part ii: Handling constraints and extending to an adaptive approach}.{\BBCQ}
\newblock
\APACjournalVolNumPages{IEEE Transactions on Evolutionary Computation}{18}{4}{602--622}.
\newblock
\begin{APACrefDOI} \doi{10.1109/TEVC.2013.2281534} \end{APACrefDOI}
\PrintBackRefs{\CurrentBib}

\bibitem [\protect \citeauthoryear {%
Kapetanios%
, Mitchell%
, Price%
\BCBL {}\ \BBA {} Fawcett%
}{%
Kapetanios%
\ \protect \BOthers {.}}{%
{\protect \APACyear {2015}}%
}]{%
kapetanios2015generalised}
\APACinsertmetastar {%
kapetanios2015generalised}%
\begin{APACrefauthors}%
Kapetanios, G.%
, Mitchell, J.%
, Price, S.%
\BCBL {}\ \BBA {} Fawcett, N.%
\end{APACrefauthors}%
\unskip\
\newblock
\APACrefYearMonthDay{2015}{}{}.
\newblock
{\BBOQ}\APACrefatitle {Generalised density forecast combinations} {Generalised density forecast combinations}.{\BBCQ}
\newblock
\APACjournalVolNumPages{Journal of Econometrics}{188}{1}{150--165}.
\newblock
\begin{APACrefDOI} \doi{10.1016/j.jeconom.2015.02.047} \end{APACrefDOI}
\PrintBackRefs{\CurrentBib}

\bibitem [\protect \citeauthoryear {%
Kennedy%
\ \BBA {} Eberhart%
}{%
Kennedy%
\ \BBA {} Eberhart%
}{%
{\protect \APACyear {1995}}%
}]{%
kennedy1995particle}
\APACinsertmetastar {%
kennedy1995particle}%
\begin{APACrefauthors}%
Kennedy, J.%
\BCBT {}\ \BBA {} Eberhart, R.%
\end{APACrefauthors}%
\unskip\
\newblock
\APACrefYearMonthDay{1995}{}{}.
\newblock
{\BBOQ}\APACrefatitle {Particle swarm optimization} {Particle swarm optimization}.{\BBCQ}
\newblock
\APACjournalVolNumPages{Proceedings of ICNN'95 - International Conference on Neural Networks}{4}{}{1942-1948}.
\newblock
\begin{APACrefDOI} \doi{10.1109/ICNN.1995.488968} \end{APACrefDOI}
\PrintBackRefs{\CurrentBib}

\bibitem [\protect \citeauthoryear {%
Kolassa%
}{%
Kolassa%
}{%
{\protect \APACyear {2016}}%
}]{%
kolassa2016evaluating}
\APACinsertmetastar {%
kolassa2016evaluating}%
\begin{APACrefauthors}%
Kolassa, S.%
\end{APACrefauthors}%
\unskip\
\newblock
\APACrefYearMonthDay{2016}{}{}.
\newblock
{\BBOQ}\APACrefatitle {Evaluating predictive count data distributions in retail sales forecasting} {Evaluating predictive count data distributions in retail sales forecasting}.{\BBCQ}
\newblock
\APACjournalVolNumPages{International Journal of Forecasting}{32}{3}{788--803}.
\newblock
\begin{APACrefDOI} \doi{10.1016/j.ijforecast.2015.12.004} \end{APACrefDOI}
\PrintBackRefs{\CurrentBib}

\bibitem [\protect \citeauthoryear {%
Kourentzes%
\ \BBA {} Athanasopoulos%
}{%
Kourentzes%
\ \BBA {} Athanasopoulos%
}{%
{\protect \APACyear {2021}}%
}]{%
kourentzes2021elucidate}
\APACinsertmetastar {%
kourentzes2021elucidate}%
\begin{APACrefauthors}%
Kourentzes, N.%
\BCBT {}\ \BBA {} Athanasopoulos, G.%
\end{APACrefauthors}%
\unskip\
\newblock
\APACrefYearMonthDay{2021}{}{}.
\newblock
{\BBOQ}\APACrefatitle {Elucidate structure in intermittent demand series} {Elucidate structure in intermittent demand series}.{\BBCQ}
\newblock
\APACjournalVolNumPages{European Journal of Operational Research}{288}{1}{141--152}.
\newblock
\begin{APACrefDOI} \doi{10.1016/j.ejor.2020.05.046} \end{APACrefDOI}
\PrintBackRefs{\CurrentBib}

\bibitem [\protect \citeauthoryear {%
Kourentzes%
, Trapero%
\BCBL {}\ \BBA {} Barrow%
}{%
Kourentzes%
\ \protect \BOthers {.}}{%
{\protect \APACyear {2020}}%
}]{%
kourentzes2020optimising}
\APACinsertmetastar {%
kourentzes2020optimising}%
\begin{APACrefauthors}%
Kourentzes, N.%
, Trapero, J\BPBI R.%
\BCBL {}\ \BBA {} Barrow, D\BPBI K.%
\end{APACrefauthors}%
\unskip\
\newblock
\APACrefYearMonthDay{2020}{}{}.
\newblock
{\BBOQ}\APACrefatitle {Optimising forecasting models for inventory planning} {Optimising forecasting models for inventory planning}.{\BBCQ}
\newblock
\APACjournalVolNumPages{International Journal of Production Economics}{225}{}{107597}.
\newblock
\begin{APACrefDOI} \doi{10.1016/j.ijpe.2019.107597} \end{APACrefDOI}
\PrintBackRefs{\CurrentBib}

\bibitem [\protect \citeauthoryear {%
Li%
, Kang%
\BCBL {}\ \BBA {} Li%
}{%
Li%
\ \protect \BOthers {.}}{%
{\protect \APACyear {2023}}%
}]{%
li2022bayesian}
\APACinsertmetastar {%
li2022bayesian}%
\begin{APACrefauthors}%
Li, L.%
, Kang, Y.%
\BCBL {}\ \BBA {} Li, F.%
\end{APACrefauthors}%
\unskip\
\newblock
\APACrefYearMonthDay{2023}{}{}.
\newblock
{\BBOQ}\APACrefatitle {Bayesian forecast combination using time-varying features} {Bayesian forecast combination using time-varying features}.{\BBCQ}
\newblock
\APACjournalVolNumPages{International Journal of Forecasting}{39}{3}{1287--1302}.
\newblock
\begin{APACrefDOI} \doi{10.1016/j.ijforecast.2022.06.002} \end{APACrefDOI}
\PrintBackRefs{\CurrentBib}

\bibitem [\protect \citeauthoryear {%
Lichtendahl~Jr%
, Grushka-Cockayne%
\BCBL {}\ \BBA {} Winkler%
}{%
Lichtendahl~Jr%
\ \protect \BOthers {.}}{%
{\protect \APACyear {2013}}%
}]{%
lichtendahl2013better}
\APACinsertmetastar {%
lichtendahl2013better}%
\begin{APACrefauthors}%
Lichtendahl~Jr, K\BPBI C.%
, Grushka-Cockayne, Y.%
\BCBL {}\ \BBA {} Winkler, R\BPBI L.%
\end{APACrefauthors}%
\unskip\
\newblock
\APACrefYearMonthDay{2013}{}{}.
\newblock
{\BBOQ}\APACrefatitle {Is it better to average probabilities or quantiles?} {Is it better to average probabilities or quantiles?}{\BBCQ}
\newblock
\APACjournalVolNumPages{Management Science}{59}{7}{1594--1611}.
\newblock
\begin{APACrefDOI} \doi{10.1287/mnsc.1120.1667} \end{APACrefDOI}
\PrintBackRefs{\CurrentBib}

\bibitem [\protect \citeauthoryear {%
Lin%
, Chen%
, Li%
\BCBL {}\ \BBA {} Shen%
}{%
Lin%
\ \protect \BOthers {.}}{%
{\protect \APACyear {2022}}%
}]{%
lin2022data}
\APACinsertmetastar {%
lin2022data}%
\begin{APACrefauthors}%
Lin, S.%
, Chen, Y.%
, Li, Y.%
\BCBL {}\ \BBA {} Shen, Z\BHBI J\BPBI M.%
\end{APACrefauthors}%
\unskip\
\newblock
\APACrefYearMonthDay{2022}{}{}.
\newblock
{\BBOQ}\APACrefatitle {Data-driven newsvendor problems regularized by a profit risk constraint} {Data-driven newsvendor problems regularized by a profit risk constraint}.{\BBCQ}
\newblock
\APACjournalVolNumPages{Production and Operations Management}{31}{4}{1630--1644}.
\newblock
\begin{APACrefDOI} \doi{10.1111/poms.13635} \end{APACrefDOI}
\PrintBackRefs{\CurrentBib}

\bibitem [\protect \citeauthoryear {%
Makridakis%
, Spiliotis%
\BCBL {}\ \BBA {} Assimakopoulos%
}{%
Makridakis%
\ \protect \BOthers {.}}{%
{\protect \APACyear {2022}}%
}]{%
makridakis2022m5}
\APACinsertmetastar {%
makridakis2022m5}%
\begin{APACrefauthors}%
Makridakis, S.%
, Spiliotis, E.%
\BCBL {}\ \BBA {} Assimakopoulos, V.%
\end{APACrefauthors}%
\unskip\
\newblock
\APACrefYearMonthDay{2022}{}{}.
\newblock
{\BBOQ}\APACrefatitle {{M5 accuracy competition: Results, findings, and conclusions}} {{M5 accuracy competition: Results, findings, and conclusions}}.{\BBCQ}
\newblock
\APACjournalVolNumPages{International Journal of Forecasting}{38}{4}{1346--1364}.
\newblock
\begin{APACrefDOI} \doi{10.1016/j.ijforecast.2021.11.013} \end{APACrefDOI}
\PrintBackRefs{\CurrentBib}

\bibitem [\protect \citeauthoryear {%
McAlinn%
, Aastveit%
, Nakajima%
\BCBL {}\ \BBA {} West%
}{%
McAlinn%
\ \protect \BOthers {.}}{%
{\protect \APACyear {2020}}%
}]{%
mcalinn2020multivariate}
\APACinsertmetastar {%
mcalinn2020multivariate}%
\begin{APACrefauthors}%
McAlinn, K.%
, Aastveit, K\BPBI A.%
, Nakajima, J.%
\BCBL {}\ \BBA {} West, M.%
\end{APACrefauthors}%
\unskip\
\newblock
\APACrefYearMonthDay{2020}{}{}.
\newblock
{\BBOQ}\APACrefatitle {Multivariate Bayesian predictive synthesis in macroeconomic forecasting} {Multivariate bayesian predictive synthesis in macroeconomic forecasting}.{\BBCQ}
\newblock
\APACjournalVolNumPages{Journal of the American Statistical Association}{115}{531}{1092--1110}.
\newblock
\begin{APACrefDOI} \doi{10.1080/01621459.2019.1660171} \end{APACrefDOI}
\PrintBackRefs{\CurrentBib}

\bibitem [\protect \citeauthoryear {%
McAlinn%
\ \BBA {} West%
}{%
McAlinn%
\ \BBA {} West%
}{%
{\protect \APACyear {2019}}%
}]{%
mcalinn2019dynamic}
\APACinsertmetastar {%
mcalinn2019dynamic}%
\begin{APACrefauthors}%
McAlinn, K.%
\BCBT {}\ \BBA {} West, M.%
\end{APACrefauthors}%
\unskip\
\newblock
\APACrefYearMonthDay{2019}{}{}.
\newblock
{\BBOQ}\APACrefatitle {Dynamic Bayesian predictive synthesis in time series forecasting} {Dynamic bayesian predictive synthesis in time series forecasting}.{\BBCQ}
\newblock
\APACjournalVolNumPages{Journal of Econometrics}{210}{1}{155--169}.
\newblock
\begin{APACrefDOI} \doi{10.1016/j.jeconom.2018.11.010} \end{APACrefDOI}
\PrintBackRefs{\CurrentBib}

\bibitem [\protect \citeauthoryear {%
O'Hagan%
\ \protect \BOthers {.}}{%
O'Hagan%
\ \protect \BOthers {.}}{%
{\protect \APACyear {2006}}%
}]{%
o2006uncertain}
\APACinsertmetastar {%
o2006uncertain}%
\begin{APACrefauthors}%
O'Hagan, A.%
, Buck, C\BPBI E.%
, Daneshkhah, A.%
, Eiser, J\BPBI R.%
, Garthwaite, P\BPBI H.%
, Jenkinson, D\BPBI J.%
\BDBL {}Rakow, T.%
\end{APACrefauthors}%
\unskip\
\newblock
\APACrefYear{2006}.
\newblock
\APACrefbtitle {Uncertain judgements: eliciting experts' probabilities} {Uncertain judgements: eliciting experts' probabilities}.
\newblock
\APACaddressPublisher{}{John Wiley \& Sons}.
\PrintBackRefs{\CurrentBib}

\bibitem [\protect \citeauthoryear {%
Olivares-Nadal%
}{%
Olivares-Nadal%
}{%
{\protect \APACyear {2024}}%
}]{%
olivares2024constructing}
\APACinsertmetastar {%
olivares2024constructing}%
\begin{APACrefauthors}%
Olivares-Nadal, A\BPBI V.%
\end{APACrefauthors}%
\unskip\
\newblock
\APACrefYearMonthDay{2024}{}{}.
\newblock
{\BBOQ}\APACrefatitle {{Constructing decision rules for multiproduct newsvendors: An integrated estimation-and-optimization framework}} {{Constructing decision rules for multiproduct newsvendors: An integrated estimation-and-optimization framework}}.{\BBCQ}
\newblock
\APACjournalVolNumPages{European Journal of Operational Research}{315}{3}{1021--1037}.
\newblock
\begin{APACrefDOI} \doi{10.1016/j.ejor.2024.01.014} \end{APACrefDOI}
\PrintBackRefs{\CurrentBib}

\bibitem [\protect \citeauthoryear {%
Opschoor%
, Van~Dijk%
\BCBL {}\ \BBA {} van~der Wel%
}{%
Opschoor%
\ \protect \BOthers {.}}{%
{\protect \APACyear {2017}}%
}]{%
opschoor2017combining}
\APACinsertmetastar {%
opschoor2017combining}%
\begin{APACrefauthors}%
Opschoor, A.%
, Van~Dijk, D.%
\BCBL {}\ \BBA {} van~der Wel, M.%
\end{APACrefauthors}%
\unskip\
\newblock
\APACrefYearMonthDay{2017}{}{}.
\newblock
{\BBOQ}\APACrefatitle {Combining density forecasts using focused scoring rules} {Combining density forecasts using focused scoring rules}.{\BBCQ}
\newblock
\APACjournalVolNumPages{Journal of Applied Econometrics}{32}{7}{1298--1313}.
\newblock
\begin{APACrefDOI} \doi{10.1002/jae.2575} \end{APACrefDOI}
\PrintBackRefs{\CurrentBib}

\bibitem [\protect \citeauthoryear {%
Oroojlooyjadid%
, Snyder%
\BCBL {}\ \BBA {} Tak{\'a}{\v{c}}%
}{%
Oroojlooyjadid%
\ \protect \BOthers {.}}{%
{\protect \APACyear {2020}}%
}]{%
oroojlooyjadid2020applying}
\APACinsertmetastar {%
oroojlooyjadid2020applying}%
\begin{APACrefauthors}%
Oroojlooyjadid, A.%
, Snyder, L\BPBI V.%
\BCBL {}\ \BBA {} Tak{\'a}{\v{c}}, M.%
\end{APACrefauthors}%
\unskip\
\newblock
\APACrefYearMonthDay{2020}{}{}.
\newblock
{\BBOQ}\APACrefatitle {Applying deep learning to the newsvendor problem} {Applying deep learning to the newsvendor problem}.{\BBCQ}
\newblock
\APACjournalVolNumPages{IISE Transactions}{52}{4}{444--463}.
\newblock
\begin{APACrefDOI} \doi{10.1080/24725854.2019.1632502} \end{APACrefDOI}
\PrintBackRefs{\CurrentBib}

\bibitem [\protect \citeauthoryear {%
Petropoulos%
\ \BBA {} Kourentzes%
}{%
Petropoulos%
\ \BBA {} Kourentzes%
}{%
{\protect \APACyear {2015}}%
}]{%
petropoulos2015forecast}
\APACinsertmetastar {%
petropoulos2015forecast}%
\begin{APACrefauthors}%
Petropoulos, F.%
\BCBT {}\ \BBA {} Kourentzes, N.%
\end{APACrefauthors}%
\unskip\
\newblock
\APACrefYearMonthDay{2015}{}{}.
\newblock
{\BBOQ}\APACrefatitle {Forecast combinations for intermittent demand} {Forecast combinations for intermittent demand}.{\BBCQ}
\newblock
\APACjournalVolNumPages{Journal of the Operational Research Society}{66}{6}{914--924}.
\newblock
\begin{APACrefDOI} \doi{10.1057/jors.2014.62} \end{APACrefDOI}
\PrintBackRefs{\CurrentBib}

\bibitem [\protect \citeauthoryear {%
Qi%
\ \protect \BOthers {.}}{%
Qi%
\ \protect \BOthers {.}}{%
{\protect \APACyear {2023}}%
}]{%
qi2023practical}
\APACinsertmetastar {%
qi2023practical}%
\begin{APACrefauthors}%
Qi, M.%
, Shi, Y.%
, Qi, Y.%
, Ma, C.%
, Yuan, R.%
, Wu, D.%
\BCBL {}\ \BBA {} Shen, Z\BHBI J.%
\end{APACrefauthors}%
\unskip\
\newblock
\APACrefYearMonthDay{2023}{}{}.
\newblock
{\BBOQ}\APACrefatitle {A practical end-to-end inventory management model with deep learning} {A practical end-to-end inventory management model with deep learning}.{\BBCQ}
\newblock
\APACjournalVolNumPages{Management Science}{69}{2}{759--773}.
\newblock
\begin{APACrefDOI} \doi{10.1287/mnsc.2022.4564} \end{APACrefDOI}
\PrintBackRefs{\CurrentBib}

\bibitem [\protect \citeauthoryear {%
Raftery%
, Gneiting%
, Balabdaoui%
\BCBL {}\ \BBA {} Polakowski%
}{%
Raftery%
\ \protect \BOthers {.}}{%
{\protect \APACyear {2005}}%
}]{%
raftery2005using}
\APACinsertmetastar {%
raftery2005using}%
\begin{APACrefauthors}%
Raftery, A\BPBI E.%
, Gneiting, T.%
, Balabdaoui, F.%
\BCBL {}\ \BBA {} Polakowski, M.%
\end{APACrefauthors}%
\unskip\
\newblock
\APACrefYearMonthDay{2005}{}{}.
\newblock
{\BBOQ}\APACrefatitle {{Using Bayesian model averaging to calibrate forecast ensembles}} {{Using Bayesian model averaging to calibrate forecast ensembles}}.{\BBCQ}
\newblock
\APACjournalVolNumPages{Monthly Weather Review}{133}{5}{1155--1174}.
\newblock
\begin{APACrefDOI} \doi{10.1175/MWR2906.1} \end{APACrefDOI}
\PrintBackRefs{\CurrentBib}

\bibitem [\protect \citeauthoryear {%
Ranjan%
\ \BBA {} Gneiting%
}{%
Ranjan%
\ \BBA {} Gneiting%
}{%
{\protect \APACyear {2010}}%
}]{%
ranjan2010combining}
\APACinsertmetastar {%
ranjan2010combining}%
\begin{APACrefauthors}%
Ranjan, R.%
\BCBT {}\ \BBA {} Gneiting, T.%
\end{APACrefauthors}%
\unskip\
\newblock
\APACrefYearMonthDay{2010}{}{}.
\newblock
{\BBOQ}\APACrefatitle {Combining probability forecasts} {Combining probability forecasts}.{\BBCQ}
\newblock
\APACjournalVolNumPages{Journal of the Royal Statistical Society: Series B (Statistical Methodology)}{72}{1}{71--91}.
\newblock
\begin{APACrefDOI} \doi{10.1111/j.1467-9868.2009.00726.x} \end{APACrefDOI}
\PrintBackRefs{\CurrentBib}

\bibitem [\protect \citeauthoryear {%
Sadana%
\ \protect \BOthers {.}}{%
Sadana%
\ \protect \BOthers {.}}{%
{\protect \APACyear {2025}}%
}]{%
sadana2025survey}
\APACinsertmetastar {%
sadana2025survey}%
\begin{APACrefauthors}%
Sadana, U.%
, Chenreddy, A.%
, Delage, E.%
, Forel, A.%
, Frejinger, E.%
\BCBL {}\ \BBA {} Vidal, T.%
\end{APACrefauthors}%
\unskip\
\newblock
\APACrefYearMonthDay{2025}{}{}.
\newblock
{\BBOQ}\APACrefatitle {A survey of contextual optimization methods for decision-making under uncertainty} {A survey of contextual optimization methods for decision-making under uncertainty}.{\BBCQ}
\newblock
\APACjournalVolNumPages{European Journal of Operational Research}{320}{2}{271--289}.
\newblock
\begin{APACrefDOI} \doi{10.1016/j.ejor.2024.03.020} \end{APACrefDOI}
\PrintBackRefs{\CurrentBib}

\bibitem [\protect \citeauthoryear {%
Snyder%
, Ord%
\BCBL {}\ \BBA {} Beaumont%
}{%
Snyder%
\ \protect \BOthers {.}}{%
{\protect \APACyear {2012}}%
}]{%
snyder2012forecasting}
\APACinsertmetastar {%
snyder2012forecasting}%
\begin{APACrefauthors}%
Snyder, R\BPBI D.%
, Ord, J\BPBI K.%
\BCBL {}\ \BBA {} Beaumont, A.%
\end{APACrefauthors}%
\unskip\
\newblock
\APACrefYearMonthDay{2012}{}{}.
\newblock
{\BBOQ}\APACrefatitle {{Forecasting the intermittent demand for slow-moving inventories: A modelling approach}} {{Forecasting the intermittent demand for slow-moving inventories: A modelling approach}}.{\BBCQ}
\newblock
\APACjournalVolNumPages{International Journal of Forecasting}{28}{2}{485--496}.
\newblock
\begin{APACrefDOI} \doi{10.1016/j.ijforecast.2011.03.009} \end{APACrefDOI}
\PrintBackRefs{\CurrentBib}

\bibitem [\protect \citeauthoryear {%
Spiliotis%
, Makridakis%
, Kaltsounis%
\BCBL {}\ \BBA {} Assimakopoulos%
}{%
Spiliotis%
\ \protect \BOthers {.}}{%
{\protect \APACyear {2021}}%
}]{%
SPILIOTIS2021108237}
\APACinsertmetastar {%
SPILIOTIS2021108237}%
\begin{APACrefauthors}%
Spiliotis, E.%
, Makridakis, S.%
, Kaltsounis, A.%
\BCBL {}\ \BBA {} Assimakopoulos, V.%
\end{APACrefauthors}%
\unskip\
\newblock
\APACrefYearMonthDay{2021}{}{}.
\newblock
{\BBOQ}\APACrefatitle {{Product sales probabilistic forecasting: An empirical evaluation using the M5 competition data}} {{Product sales probabilistic forecasting: An empirical evaluation using the M5 competition data}}.{\BBCQ}
\newblock
\APACjournalVolNumPages{International Journal of Production Economics}{240}{}{108237}.
\newblock
\begin{APACrefDOI} \doi{https://doi.org/10.1016/j.ijpe.2021.108237} \end{APACrefDOI}
\PrintBackRefs{\CurrentBib}

\bibitem [\protect \citeauthoryear {%
Stone%
}{%
Stone%
}{%
{\protect \APACyear {1961}}%
}]{%
stone1961opinion}
\APACinsertmetastar {%
stone1961opinion}%
\begin{APACrefauthors}%
Stone, M.%
\end{APACrefauthors}%
\unskip\
\newblock
\APACrefYearMonthDay{1961}{}{}.
\newblock
{\BBOQ}\APACrefatitle {The Opinion Pool} {The opinion pool}.{\BBCQ}
\newblock
\APACjournalVolNumPages{The Annals of Mathematical Statistics}{32}{4}{1339--1342}.
\newblock
\begin{APACrefDOI} \doi{10.1214/aoms/1177704873} \end{APACrefDOI}
\PrintBackRefs{\CurrentBib}

\bibitem [\protect \citeauthoryear {%
Stratigakos%
, Pineda%
\BCBL {}\ \BBA {} Morales%
}{%
Stratigakos%
\ \protect \BOthers {.}}{%
{\protect \APACyear {2025}}%
}]{%
stratigakos2025decision}
\APACinsertmetastar {%
stratigakos2025decision}%
\begin{APACrefauthors}%
Stratigakos, A.%
, Pineda, S.%
\BCBL {}\ \BBA {} Morales, J\BPBI M.%
\end{APACrefauthors}%
\unskip\
\newblock
\APACrefYearMonthDay{2025}{}{}.
\newblock
{\BBOQ}\APACrefatitle {Decision-focused linear pooling for probabilistic forecast combination} {Decision-focused linear pooling for probabilistic forecast combination}.{\BBCQ}
\newblock
\APACjournalVolNumPages{International Journal of Forecasting}{41}{3}{1112--1125}.
\newblock
\begin{APACrefDOI} \doi{10.1016/j.ijforecast.2024.11.006} \end{APACrefDOI}
\PrintBackRefs{\CurrentBib}

\bibitem [\protect \citeauthoryear {%
Tallman%
\ \BBA {} West%
}{%
Tallman%
\ \BBA {} West%
}{%
{\protect \APACyear {2024}}%
}]{%
tallman2024bayesian}
\APACinsertmetastar {%
tallman2024bayesian}%
\begin{APACrefauthors}%
Tallman, E.%
\BCBT {}\ \BBA {} West, M.%
\end{APACrefauthors}%
\unskip\
\newblock
\APACrefYearMonthDay{2024}{}{}.
\newblock
{\BBOQ}\APACrefatitle {Bayesian predictive decision synthesis} {Bayesian predictive decision synthesis}.{\BBCQ}
\newblock
\APACjournalVolNumPages{Journal of the Royal Statistical Society Series B: Statistical Methodology}{86}{2}{340--363}.
\newblock
\begin{APACrefDOI} \doi{10.1093/jrsssb/qkad109} \end{APACrefDOI}
\PrintBackRefs{\CurrentBib}

\bibitem [\protect \citeauthoryear {%
Taylor%
\ \BBA {} Meng%
}{%
Taylor%
\ \BBA {} Meng%
}{%
{\protect \APACyear {2026}}%
}]{%
taylor2026angular}
\APACinsertmetastar {%
taylor2026angular}%
\begin{APACrefauthors}%
Taylor, J\BPBI W.%
\BCBT {}\ \BBA {} Meng, X.%
\end{APACrefauthors}%
\unskip\
\newblock
\APACrefYearMonthDay{2026}{}{}.
\newblock
{\BBOQ}\APACrefatitle {Angular combining of forecasts of probability distributions} {Angular combining of forecasts of probability distributions}.{\BBCQ}
\newblock
\APACjournalVolNumPages{Management Science}{72}{3}{2111--2133}.
\newblock
\begin{APACrefDOI} \doi{10.1287/mnsc.2024.05558} \end{APACrefDOI}
\PrintBackRefs{\CurrentBib}

\bibitem [\protect \citeauthoryear {%
Teunter%
\ \BBA {} Duncan%
}{%
Teunter%
\ \BBA {} Duncan%
}{%
{\protect \APACyear {2009}}%
}]{%
teunter2009forecasting}
\APACinsertmetastar {%
teunter2009forecasting}%
\begin{APACrefauthors}%
Teunter, R\BPBI H.%
\BCBT {}\ \BBA {} Duncan, L.%
\end{APACrefauthors}%
\unskip\
\newblock
\APACrefYearMonthDay{2009}{}{}.
\newblock
{\BBOQ}\APACrefatitle {Forecasting intermittent demand: a comparative study} {Forecasting intermittent demand: a comparative study}.{\BBCQ}
\newblock
\APACjournalVolNumPages{Journal of the Operational Research Society}{60}{3}{321--329}.
\newblock
\begin{APACrefDOI} \doi{10.1057/palgrave.jors.2602569} \end{APACrefDOI}
\PrintBackRefs{\CurrentBib}

\bibitem [\protect \citeauthoryear {%
Theodorou%
, Spiliotis%
\BCBL {}\ \BBA {} Assimakopoulos%
}{%
Theodorou%
\ \protect \BOthers {.}}{%
{\protect \APACyear {2025}}%
}]{%
THEODOROU2025414}
\APACinsertmetastar {%
THEODOROU2025414}%
\begin{APACrefauthors}%
Theodorou, E.%
, Spiliotis, E.%
\BCBL {}\ \BBA {} Assimakopoulos, V.%
\end{APACrefauthors}%
\unskip\
\newblock
\APACrefYearMonthDay{2025}{}{}.
\newblock
{\BBOQ}\APACrefatitle {{Forecast accuracy and inventory performance: Insights on their relationship from the M5 competition data}} {{Forecast accuracy and inventory performance: Insights on their relationship from the M5 competition data}}.{\BBCQ}
\newblock
\APACjournalVolNumPages{European Journal of Operational Research}{322}{2}{414-426}.
\newblock
\begin{APACrefDOI} \doi{https://doi.org/10.1016/j.ejor.2024.12.033} \end{APACrefDOI}
\PrintBackRefs{\CurrentBib}

\bibitem [\protect \citeauthoryear {%
Thorey%
, Chaussin%
\BCBL {}\ \BBA {} Mallet%
}{%
Thorey%
\ \protect \BOthers {.}}{%
{\protect \APACyear {2018}}%
}]{%
thorey2018ensemble}
\APACinsertmetastar {%
thorey2018ensemble}%
\begin{APACrefauthors}%
Thorey, J.%
, Chaussin, C.%
\BCBL {}\ \BBA {} Mallet, V.%
\end{APACrefauthors}%
\unskip\
\newblock
\APACrefYearMonthDay{2018}{}{}.
\newblock
{\BBOQ}\APACrefatitle {Ensemble forecast of photovoltaic power with online {CRPS} learning} {Ensemble forecast of photovoltaic power with online {CRPS} learning}.{\BBCQ}
\newblock
\APACjournalVolNumPages{International Journal of Forecasting}{34}{4}{762--773}.
\newblock
\begin{APACrefDOI} \doi{10.1016/j.ijforecast.2018.05.007} \end{APACrefDOI}
\PrintBackRefs{\CurrentBib}

\bibitem [\protect \citeauthoryear {%
Thorey%
, Mallet%
\BCBL {}\ \BBA {} Baudin%
}{%
Thorey%
\ \protect \BOthers {.}}{%
{\protect \APACyear {2017}}%
}]{%
thorey2017online}
\APACinsertmetastar {%
thorey2017online}%
\begin{APACrefauthors}%
Thorey, J.%
, Mallet, V.%
\BCBL {}\ \BBA {} Baudin, P.%
\end{APACrefauthors}%
\unskip\
\newblock
\APACrefYearMonthDay{2017}{}{}.
\newblock
{\BBOQ}\APACrefatitle {Online learning with the continuous ranked probability score for ensemble forecasting} {Online learning with the continuous ranked probability score for ensemble forecasting}.{\BBCQ}
\newblock
\APACjournalVolNumPages{Quarterly Journal of the Royal Meteorological Society}{143}{702}{521--529}.
\newblock
\begin{APACrefDOI} \doi{10.1002/qj.2940} \end{APACrefDOI}
\PrintBackRefs{\CurrentBib}

\bibitem [\protect \citeauthoryear {%
Trapero%
, Card{\'o}s%
\BCBL {}\ \BBA {} Kourentzes%
}{%
Trapero%
\ \protect \BOthers {.}}{%
{\protect \APACyear {2019}}%
}]{%
trapero2019quantile}
\APACinsertmetastar {%
trapero2019quantile}%
\begin{APACrefauthors}%
Trapero, J\BPBI R.%
, Card{\'o}s, M.%
\BCBL {}\ \BBA {} Kourentzes, N.%
\end{APACrefauthors}%
\unskip\
\newblock
\APACrefYearMonthDay{2019}{}{}.
\newblock
{\BBOQ}\APACrefatitle {Quantile forecast optimal combination to enhance safety stock estimation} {Quantile forecast optimal combination to enhance safety stock estimation}.{\BBCQ}
\newblock
\APACjournalVolNumPages{International Journal of Forecasting}{35}{1}{239--250}.
\newblock
\begin{APACrefDOI} \doi{10.1016/j.ijforecast.2018.05.009} \end{APACrefDOI}
\PrintBackRefs{\CurrentBib}

\bibitem [\protect \citeauthoryear {%
Tsai%
\ \BBA {} Chen%
}{%
Tsai%
\ \BBA {} Chen%
}{%
{\protect \APACyear {2017}}%
}]{%
tsai2017simulation}
\APACinsertmetastar {%
tsai2017simulation}%
\begin{APACrefauthors}%
Tsai, S\BPBI C.%
\BCBT {}\ \BBA {} Chen, S\BPBI T.%
\end{APACrefauthors}%
\unskip\
\newblock
\APACrefYearMonthDay{2017}{}{}.
\newblock
{\BBOQ}\APACrefatitle {A simulation-based multi-objective optimization framework: A case study on inventory management} {A simulation-based multi-objective optimization framework: A case study on inventory management}.{\BBCQ}
\newblock
\APACjournalVolNumPages{Omega}{70}{}{148--159}.
\newblock
\begin{APACrefDOI} \doi{10.1016/j.omega.2016.09.007} \end{APACrefDOI}
\PrintBackRefs{\CurrentBib}

\bibitem [\protect \citeauthoryear {%
Van~der Meer%
, Pinson%
, Camal%
\BCBL {}\ \BBA {} Kariniotakis%
}{%
Van~der Meer%
\ \protect \BOthers {.}}{%
{\protect \APACyear {2024}}%
}]{%
van2024crps}
\APACinsertmetastar {%
van2024crps}%
\begin{APACrefauthors}%
Van~der Meer, D.%
, Pinson, P.%
, Camal, S.%
\BCBL {}\ \BBA {} Kariniotakis, G.%
\end{APACrefauthors}%
\unskip\
\newblock
\APACrefYearMonthDay{2024}{}{}.
\newblock
{\BBOQ}\APACrefatitle {{CRPS}-based online learning for nonlinear probabilistic forecast combination} {{CRPS}-based online learning for nonlinear probabilistic forecast combination}.{\BBCQ}
\newblock
\APACjournalVolNumPages{International Journal of Forecasting}{40}{4}{1449--1466}.
\newblock
\begin{APACrefDOI} \doi{10.1016/j.ijforecast.2023.12.005} \end{APACrefDOI}
\PrintBackRefs{\CurrentBib}

\bibitem [\protect \citeauthoryear {%
Wallis%
}{%
Wallis%
}{%
{\protect \APACyear {2005}}%
}]{%
wallis2005combining}
\APACinsertmetastar {%
wallis2005combining}%
\begin{APACrefauthors}%
Wallis, K\BPBI F.%
\end{APACrefauthors}%
\unskip\
\newblock
\APACrefYearMonthDay{2005}{}{}.
\newblock
{\BBOQ}\APACrefatitle {Combining density and interval forecasts: a modest proposal} {Combining density and interval forecasts: a modest proposal}.{\BBCQ}
\newblock
\APACjournalVolNumPages{Oxford Bulletin of Economics and Statistics}{67}{}{983--994}.
\newblock
\begin{APACrefDOI} \doi{10.1111/j.1468-0084.2005.00148.x} \end{APACrefDOI}
\PrintBackRefs{\CurrentBib}

\bibitem [\protect \citeauthoryear {%
J.~Wang%
, An%
, Li%
\BCBL {}\ \BBA {} Lu%
}{%
J.~Wang%
\ \protect \BOthers {.}}{%
{\protect \APACyear {2022}}%
}]{%
wang2022novel}
\APACinsertmetastar {%
wang2022novel}%
\begin{APACrefauthors}%
Wang, J.%
, An, Y.%
, Li, Z.%
\BCBL {}\ \BBA {} Lu, H.%
\end{APACrefauthors}%
\unskip\
\newblock
\APACrefYearMonthDay{2022}{}{}.
\newblock
{\BBOQ}\APACrefatitle {A novel combined forecasting model based on neural networks, deep learning approaches, and multi-objective optimization for short-term wind speed forecasting} {A novel combined forecasting model based on neural networks, deep learning approaches, and multi-objective optimization for short-term wind speed forecasting}.{\BBCQ}
\newblock
\APACjournalVolNumPages{Energy}{251}{}{123960}.
\newblock
\begin{APACrefDOI} \doi{10.1016/j.energy.2022.123960} \end{APACrefDOI}
\PrintBackRefs{\CurrentBib}

\bibitem [\protect \citeauthoryear {%
S.~Wang%
, Kang%
\BCBL {}\ \BBA {} Petropoulos%
}{%
S.~Wang%
\ \protect \BOthers {.}}{%
{\protect \APACyear {2024}}%
}]{%
wang2024combining}
\APACinsertmetastar {%
wang2024combining}%
\begin{APACrefauthors}%
Wang, S.%
, Kang, Y.%
\BCBL {}\ \BBA {} Petropoulos, F.%
\end{APACrefauthors}%
\unskip\
\newblock
\APACrefYearMonthDay{2024}{}{}.
\newblock
{\BBOQ}\APACrefatitle {Combining probabilistic forecasts of intermittent demand} {Combining probabilistic forecasts of intermittent demand}.{\BBCQ}
\newblock
\APACjournalVolNumPages{European Journal of Operational Research}{315}{3}{1038--1048}.
\newblock
\begin{APACrefDOI} \doi{10.1016/j.ejor.2024.01.032} \end{APACrefDOI}
\PrintBackRefs{\CurrentBib}

\bibitem [\protect \citeauthoryear {%
X.~Wang%
, Hyndman%
, Li%
\BCBL {}\ \BBA {} Kang%
}{%
X.~Wang%
\ \protect \BOthers {.}}{%
{\protect \APACyear {2023}}%
}]{%
wang2023forecast}
\APACinsertmetastar {%
wang2023forecast}%
\begin{APACrefauthors}%
Wang, X.%
, Hyndman, R\BPBI J.%
, Li, F.%
\BCBL {}\ \BBA {} Kang, Y.%
\end{APACrefauthors}%
\unskip\
\newblock
\APACrefYearMonthDay{2023}{}{}.
\newblock
{\BBOQ}\APACrefatitle {{Forecast combinations: An over 50-year review}} {{Forecast combinations: An over 50-year review}}.{\BBCQ}
\newblock
\APACjournalVolNumPages{International Journal of Forecasting}{39}{4}{1518--1547}.
\newblock
\begin{APACrefDOI} \doi{10.1016/j.ijforecast.2022.11.005} \end{APACrefDOI}
\PrintBackRefs{\CurrentBib}

\bibitem [\protect \citeauthoryear {%
Waychal%
, Laha%
\BCBL {}\ \BBA {} Sinha%
}{%
Waychal%
\ \protect \BOthers {.}}{%
{\protect \APACyear {2024}}%
}]{%
waychal2024adaptive}
\APACinsertmetastar {%
waychal2024adaptive}%
\begin{APACrefauthors}%
Waychal, N.%
, Laha, A\BPBI K.%
\BCBL {}\ \BBA {} Sinha, A.%
\end{APACrefauthors}%
\unskip\
\newblock
\APACrefYearMonthDay{2024}{}{}.
\newblock
{\BBOQ}\APACrefatitle {An adaptive multi-objective optimal forecast combination and its application for predicting intermittent demand} {An adaptive multi-objective optimal forecast combination and its application for predicting intermittent demand}.{\BBCQ}
\newblock
\APACjournalVolNumPages{Journal of the Operational Research Society}{75}{9}{1813--1825}.
\newblock
\begin{APACrefDOI} \doi{10.1080/01605682.2023.2277865} \end{APACrefDOI}
\PrintBackRefs{\CurrentBib}

\bibitem [\protect \citeauthoryear {%
Weber%
}{%
Weber%
}{%
{\protect \APACyear {2025}}%
}]{%
weber2025relatively}
\APACinsertmetastar {%
weber2025relatively}%
\begin{APACrefauthors}%
Weber, T\BPBI A.%
\end{APACrefauthors}%
\unskip\
\newblock
\APACrefYearMonthDay{2025}{}{}.
\newblock
{\BBOQ}\APACrefatitle {Relatively Robust Multicriteria Decisions} {Relatively robust multicriteria decisions}.{\BBCQ}
\newblock
\APACjournalVolNumPages{Management Science}{0}{}{0}.
\newblock
\begin{APACrefDOI} \doi{10.1287/mnsc.2025.00510} \end{APACrefDOI}
\PrintBackRefs{\CurrentBib}

\bibitem [\protect \citeauthoryear {%
Willemain%
, Smart%
\BCBL {}\ \BBA {} Schwarz%
}{%
Willemain%
\ \protect \BOthers {.}}{%
{\protect \APACyear {2004}}%
}]{%
willemain2004new}
\APACinsertmetastar {%
willemain2004new}%
\begin{APACrefauthors}%
Willemain, T\BPBI R.%
, Smart, C\BPBI N.%
\BCBL {}\ \BBA {} Schwarz, H\BPBI F.%
\end{APACrefauthors}%
\unskip\
\newblock
\APACrefYearMonthDay{2004}{}{}.
\newblock
{\BBOQ}\APACrefatitle {A new approach to forecasting intermittent demand for service parts inventories} {A new approach to forecasting intermittent demand for service parts inventories}.{\BBCQ}
\newblock
\APACjournalVolNumPages{International Journal of Forecasting}{20}{3}{375-387}.
\newblock
\begin{APACrefDOI} \doi{https://doi.org/10.1016/S0169-2070(03)00013-X} \end{APACrefDOI}
\PrintBackRefs{\CurrentBib}

\bibitem [\protect \citeauthoryear {%
Wright%
}{%
Wright%
}{%
{\protect \APACyear {2008}}%
}]{%
wright2008bayesian}
\APACinsertmetastar {%
wright2008bayesian}%
\begin{APACrefauthors}%
Wright, J\BPBI H.%
\end{APACrefauthors}%
\unskip\
\newblock
\APACrefYearMonthDay{2008}{}{}.
\newblock
{\BBOQ}\APACrefatitle {Bayesian model averaging and exchange rate forecasts} {Bayesian model averaging and exchange rate forecasts}.{\BBCQ}
\newblock
\APACjournalVolNumPages{Journal of Econometrics}{146}{2}{329--341}.
\newblock
\begin{APACrefDOI} \doi{10.1016/j.jeconom.2008.08.012} \end{APACrefDOI}
\PrintBackRefs{\CurrentBib}

\bibitem [\protect \citeauthoryear {%
Xing%
, Huang%
, Wang%
\BCBL {}\ \BBA {} Wang%
}{%
Xing%
\ \protect \BOthers {.}}{%
{\protect \APACyear {2024}}%
}]{%
xing2024novel}
\APACinsertmetastar {%
xing2024novel}%
\begin{APACrefauthors}%
Xing, Q.%
, Huang, X.%
, Wang, J.%
\BCBL {}\ \BBA {} Wang, S.%
\end{APACrefauthors}%
\unskip\
\newblock
\APACrefYearMonthDay{2024}{}{}.
\newblock
{\BBOQ}\APACrefatitle {A novel multivariate combined power load forecasting system based on feature selection and multi-objective intelligent optimization} {A novel multivariate combined power load forecasting system based on feature selection and multi-objective intelligent optimization}.{\BBCQ}
\newblock
\APACjournalVolNumPages{Expert Systems with Applications}{244}{}{122970}.
\newblock
\begin{APACrefDOI} \doi{10.1016/j.eswa.2023.122970} \end{APACrefDOI}
\PrintBackRefs{\CurrentBib}

\bibitem [\protect \citeauthoryear {%
Yang%
, Zang%
, Wu%
\BCBL {}\ \BBA {} Hao%
}{%
Yang%
\ \protect \BOthers {.}}{%
{\protect \APACyear {2024}}%
}]{%
yang2024new}
\APACinsertmetastar {%
yang2024new}%
\begin{APACrefauthors}%
Yang, W.%
, Zang, X.%
, Wu, C.%
\BCBL {}\ \BBA {} Hao, Y.%
\end{APACrefauthors}%
\unskip\
\newblock
\APACrefYearMonthDay{2024}{}{}.
\newblock
{\BBOQ}\APACrefatitle {{A new multi-objective ensemble wind speed forecasting system: Mixed-frequency interval-valued modeling paradigm}} {{A new multi-objective ensemble wind speed forecasting system: Mixed-frequency interval-valued modeling paradigm}}.{\BBCQ}
\newblock
\APACjournalVolNumPages{Energy}{304}{}{131963}.
\newblock
\begin{APACrefDOI} \doi{10.1016/j.energy.2024.131963} \end{APACrefDOI}
\PrintBackRefs{\CurrentBib}

\bibitem [\protect \citeauthoryear {%
Zhang%
, Yang%
\BCBL {}\ \BBA {} Gao%
}{%
Zhang%
\ \protect \BOthers {.}}{%
{\protect \APACyear {2024}}%
}]{%
zhang2024optimal}
\APACinsertmetastar {%
zhang2024optimal}%
\begin{APACrefauthors}%
Zhang, L.%
, Yang, J.%
\BCBL {}\ \BBA {} Gao, R.%
\end{APACrefauthors}%
\unskip\
\newblock
\APACrefYearMonthDay{2024}{}{}.
\newblock
{\BBOQ}\APACrefatitle {Optimal robust policy for feature-based newsvendor} {Optimal robust policy for feature-based newsvendor}.{\BBCQ}
\newblock
\APACjournalVolNumPages{Management Science}{70}{4}{2315--2329}.
\newblock
\begin{APACrefDOI} \doi{10.1287/mnsc.2023.4810} \end{APACrefDOI}
\PrintBackRefs{\CurrentBib}

\bibitem [\protect \citeauthoryear {%
Zhou%
\ \BBA {} Viswanathan%
}{%
Zhou%
\ \BBA {} Viswanathan%
}{%
{\protect \APACyear {2011}}%
}]{%
zhou2011comparison}
\APACinsertmetastar {%
zhou2011comparison}%
\begin{APACrefauthors}%
Zhou, C.%
\BCBT {}\ \BBA {} Viswanathan, S.%
\end{APACrefauthors}%
\unskip\
\newblock
\APACrefYearMonthDay{2011}{}{}.
\newblock
{\BBOQ}\APACrefatitle {Comparison of a new bootstrapping method with parametric approaches for safety stock determination in service parts inventory systems} {Comparison of a new bootstrapping method with parametric approaches for safety stock determination in service parts inventory systems}.{\BBCQ}
\newblock
\APACjournalVolNumPages{International Journal of Production Economics}{133}{1}{481--485}.
\newblock
\begin{APACrefDOI} \doi{10.1016/j.ijpe.2010.09.021} \end{APACrefDOI}
\PrintBackRefs{\CurrentBib}

\end{thebibliography}

\appendix

\section{Comparison of selection methods}

As discussed in Section 4.2, MOO yields a set of Pareto-optimal solutions rather than a single global optimum. Consequently, a decision-making strategy is required to select a final solution from the Pareto frontier. While the results in the Section 5.2 utilize the ideal-point method, this appendix compares its performance against the performance-index method. Additionally, we include the average of the Pareto set as a benchmark, representing a baseline or random selection strategy.

Tables \ref{tab:selection-80}, \ref{tab:selection-90}, and \ref{tab:selection-95} present the comparative performance of these selection strategies. The values in brackets indicate the percentage point deviation relative to the average performance of the Pareto set, with the percent sign omitted; negative values denote that the selection method outperforms the benchmark (i.e., achieves lower error or cost).

The results indicate that the choice of selection method materially impacts performance. For NSGA-\Romannum{3}-hs, the ideal-point method proves superior to the performance-index method in 9 out of 12 evaluation metrics. However, for NSGA-\Romannum{3}-c, the results are more mixed, with the ideal-point method outperforming the performance-index method in only 5 cases.

When compared against the baseline average, the ideal-point method demonstrates consistent superiority across both MOO algorithms. In contrast, the performance-index method shows instability; while it provides an advantage in NSGA-\Romannum{3}-c, it performs worse than the simple average in NSGA-\Romannum{3}-hs. These findings suggest that the ideal-point method offers a more robust and reliable strategy for solution selection. Therefore, throughout the main text, all forecasts derived from MOO-based combinations utilize the ideal-point selection method.

\begin{table}[!ht]
\centering
\caption{Forecast and decision metrics of different selection methods with cost parameters $(c_1,c_2)=(1,4)$ for M5 data. The values in brackets are the percentage point deviation relative to the average performance of the Pareto set, omitting the percent sign.}
\small
\label{tab:selection-80}
{\begin{tabular}{llccccc}
\toprule
Type & Method & DRPS & Cost & Holding & Stockout \\
\midrule 
\multirow{2}*{Ideal point} & NSGA-\Romannum{3}-c & 1.1419(0.08) & 65.3958(-0.10) & 33.5079(-0.57)&  7.9720(0.39) \\
~ & NSGA-\Romannum{3}-hs  & 1.1427(-1.21) & 65.7930(-0.52) & 33.5842(1.08)& 8.0522(-2.12) \\
\hline
\multirow{2}*{Performance index} & NSGA-\Romannum{3}-c & 1.1395(-0.14) & 65.2841(-0.27)& 33.3426(-1.06)& 7.9854(0.56)\\
~ &  NSGA-\Romannum{3}-hs  & 1.1627(0.52) & 66.5067(0.56)& 33.5089(0.85)&  8.2495(0.27)\\
\hline
\multirow{2}*{Average} & NSGA-\Romannum{3}-c & 1.1410& 65.4634& 33.7008 & 7.9406\\
~ &  NSGA-\Romannum{3}-hs  & 1.1567 & 66.1349 & 33.2268&  8.2270\\
\bottomrule 
\end{tabular}}
\end{table}

\begin{table}[!ht]
\centering
\caption{Forecast and decision metrics of different selection methods with cost parameters $(c_1,c_2)=(1,9)$ for M5 data. The values in brackets are the percentage point deviation relative to the average performance of the Pareto set, omitting the percent sign.}
\small
\label{tab:selection-90}
{\begin{tabular}{llccccc}
\toprule
Type & Method & DRPS & Cost & Holding & Stockout \\
\midrule 
\multirow{2}*{Ideal point} & NSGA-\Romannum{3}-c & 1.1337(0.02) & 92.3667(-0.20) & 55.6951(-0.07)&  4.0746(-0.39) \\
~ & NSGA-\Romannum{3}-hs  & 1.1263(-1.20) & 92.4628(-0.63) & 54.9600(-0.75)& 4.1670(-0.46) \\
\hline
\multirow{2}*{Performance index} & NSGA-\Romannum{3}-c & 1.1352(0.16)& 92.1292(-0.45)& 55.1243(-1.09)&  4.1116(0.51)\\
~ &  NSGA-\Romannum{3}-hs  & 1.1437(0.32) & 93.3932(0.37)& 58.0555(4.84)&  3.9264(-6.21)\\
\hline
\multirow{2}*{Average} & NSGA-\Romannum{3}-c & 1.1334&92.5502& 55.7336 & 4.0907\\
~ &  NSGA-\Romannum{3}-hs  & 1.1400 & 93.0522 & 55.3766&  4.1862\\
\bottomrule 
\end{tabular}}
\end{table}

\begin{table}[!ht]
\centering
\caption{Forecast and decision metrics of different selection methods with cost parameters $(c_1,c_2)=(1,19)$ for M5 data. The values in brackets are the percentage point deviation relative to the average performance of the Pareto set, omitting the percent sign.}
\small
\label{tab:selection-95}
{\begin{tabular}{llccccc}
\toprule
Type & Method & DRPS & Cost & Holding & Stockout \\
\midrule 
\multirow{2}*{Ideal point} & NSGA-\Romannum{3}-c & 1.1351(-0.07) & 122.2545(-0.09) & 79.3229(0.05)&  2.2596(-1.46) \\
~ & NSGA-\Romannum{3}-hs  & 1.1230(-0.97) & 121.8918(-0.33) & 76.0481(-1.36)& 2.4128(0.32) \\
\hline
\multirow{2}*{Performance index} & NSGA-\Romannum{3}-c & 1.1400(0.36) & 121.3837(-0.80)& 78.3310(-1.20)&  2.2918(-0.05)\\
~ &  NSGA-\Romannum{3}-hs  & 1.1375(0.31) & 122.1444(-0.12)& 81.5220(5.74)&  2.1632(-10.05)\\
\hline
\multirow{2}*{Average} & NSGA-\Romannum{3}-c & 1.1360& 122.3605& 79.2838 & 2.2929\\
~ &  NSGA-\Romannum{3}-hs  & 1.1340 & 122.2929 & 77.1002&  2.4050\\
\bottomrule 
\end{tabular}}
\end{table}

\end{document}